\documentclass[a4paper,12pt]{article}
\pdfoutput=1
\renewcommand{\today}{\ifnum\number\day<10 0\fi \number\day \space%
\ifcase \month \or January\or February\or March\or April\or May%
\or June\or July\or August\or September\or October\or November\or December\fi,%
\number \year} 
\usepackage[utf8]{inputenc}
\usepackage[font=bf,skip=\baselineskip]{caption}
\usepackage{graphicx,rotating,colortbl}
\usepackage{hyperref,slashed,amsfonts}
\usepackage{color}
\usepackage{amsmath,amssymb,bbold}
\usepackage{mathtools,multirow}
\hypersetup{colorlinks,bookmarksopen,bookmarksnumbered,
linkcolor=blus,pdfstartview=FitH,urlcolor=rossos,citecolor=verde}

\def\circa#1{\,\raise.3ex\hbox{$#1$\kern-.75em\lower1ex\hbox{$\sim$}}\,}
\newcommand{\ev}[1]{\langle #1 \rangle}
\usepackage{mathtools}
\usepackage{tikzfeynman}
\usepackage{youngtab}
\usepackage{multicol}
\usepackage{color}
\definecolor{rosso}{cmyk}{0,1,1,0.4}
\definecolor{rossos}{cmyk}{0,1,1,0.55}
\definecolor{rossoc}{cmyk}{0,1,1,0.2}
\definecolor{blu}{cmyk}{1,1,0,0.3}
\definecolor{blus}{cmyk}{1,1,0,0.6}
\definecolor{bluc}{cmyk}{1,1,0,0.1}
\definecolor{verde}{cmyk}{0.92,0,0.59,0.25}
\definecolor{verdec}{cmyk}{0.92,0,0.59,0.15}
\definecolor{verdes}{cmyk}{0.92,0,0.59,0.4}

\newcommand{\Q}{\Psi}

\def\circa#1{\,\raise.3ex\hbox{$#1$\kern-.75em\lower1ex\hbox{$\sim$}}\,}

\newcommand{\beq}{\begin{equation}}
\newcommand{\eeq}{\end{equation}}

\newcommand{\bea}{\begin{eqnarray}}
\newcommand{\eea}{\end{eqnarray}}
\newcommand{\be}{\begin{equation}}
\newcommand{\ee}{\end{equation}}
\font\tenrsfs=rsfs10 at 12pt
\font\sevenrsfs=rsfs7
\font\fiversfs=rsfs5
\newfam\rsfsfam
\textfont\rsfsfam=\tenrsfs
\scriptfont\rsfsfam=\sevenrsfs
\scriptscriptfont\rsfsfam=\fiversfs
\def\mathscr#1{{\fam\rsfsfam\relax#1}}

\def\Lag{\mathscr{L}}

\def\circa#1{\,\raise.3ex\hbox{$#1$\kern-.75em\lower1ex\hbox{$\sim$}}\,}
\makeatletter

\def\hhref#1{\href{http://arxiv.org/abs/#1}{arXiv:#1}} % in bibliography
\def\art{\@ifnextchar[{\eart}{\oart}}
\def\eart[#1]#2#3#4#5#6{{\rm #2}, {\em #3 \bf #4} {\rm (#6) #5} ({\em #1})}
\def\hepart[#1]#2{{\rm #2, \hhref{#1}}}
\newcommand{\oart}[5]{{\rm #1}, {\em #2 \bf #3} {\rm (#5) #4}}

\newcounter{alphaequation}[equation]
\def\thealphaequation{\theequation\hbox to
0.6em{\hfil\alph{alphaequation}\hfil}}
\def\eqnsystem#1{
\def\@eqnnum{{\rm (\thealphaequation)}}
\def\@@eqncr{\let\@tempa\relax \ifcase\@eqcnt \def\@tempa{& & &} \or
  \def\@tempa{& &}\or \def\@tempa{&}\fi\@tempa
  \if@eqnsw\@eqnnum\refstepcounter{alphaequation}\fi
\global\@eqnswtrue\global\@eqcnt=0\cr}
\refstepcounter{equation} \let\@currentlabel\theequation \def\@tempb{#1}
\ifx\@tempb\empty\else\label{#1}\fi
\refstepcounter{alphaequation}
\let\@currentlabel\thealphaequation
\global\@eqnswtrue\global\@eqcnt=0 \tabskip\@centering\let\\=\@eqncr
$$\halign to \displaywidth\bgroup \@eqnsel\hskip\@centering
$\displaystyle\tabskip\z@{##}$&\global\@eqcnt\@ne
\hskip2\arraycolsep\hfil${##}$\hfil& \global\@eqcnt\tw@\hskip2\arraycolsep
$\displaystyle\tabskip\z@{##}$\hfil
\tabskip\@centering&\llap{##}\tabskip\z@\cr}
\def\endeqnsystem{\@@eqncr\egroup$$\global\@ignoretrue} \makeatother

\newcommand{\SO}{\,{\rm SO}}
\newcommand{\SU}{\,{\rm SU}}
\newcommand{\SP}{\,{\rm Sp}}

\newcommand{\yt}{\tilde{y}}

\begin{document}

\vspace{1cm}

\begin{center}
{\Large \bf \color{rossos}
UV complete composite Higgs models
}\\[1cm]
{\bf Alessandro Agugliaro$^{a}$, Oleg Antipin$^{b}$, Diego Becciolini$^{a}$\\ Stefania De Curtis$^{a}$, Michele Redi$^{a}$}  
\\[7mm]

{\it $^a$INFN, Sezione di Firenze, Via G. Sansone, 1; I-50019 Sesto Fiorentino, Italy}\\
{\it $^b$Rudjer Boskovic Institute, Division of Theoretical Physics, Bijenicka 54,\\ HR-10000 Zagreb, Croatia}

\vspace{0.5cm}
%\today
\vspace{0.5cm}

{\large\bf\color{blus} Abstract}
\begin{quote}
We study confining gauge theories with fermions vectorial under the SM that produce a Higgs doublet as a Nambu-Goldstone boson. 
The vacuum misalignment required to break the electro-weak symmetry is induced by an elementary Higgs doublet with Yukawa couplings
to the new fermions. The physical Higgs is a linear combination of elementary and composite Higgses while the SM fermions remain elementary.
The full theory is renormalizable and the SM Yukawa couplings are generated from the ones of the elementary Higgs
allowing to eliminate all flavour problems but with interesting effects for Electric Dipole Moments of SM particles. 
We also discuss how ideas on the relaxation of the electro-weak scale could be realised within this framework.
\end{quote}

\newpage

\thispagestyle{empty}
\end{center}
\begin{quote}
{\large\noindent\color{blus} 
}

\end{quote}
\vspace{-1.5cm}

\tableofcontents

\setcounter{footnote}{0}

\section{Introduction}\label{sec:introduction}

One of the simplest extensions of the SM is provided by new gauge theories with fermions that are in a vectorial representation of the SM group.
Similarly to QCD, when the new gauge group confines hadrons are produced. The dynamics is such that the vacuum  does not break 
the SM symmetries and for this reason the phenomenology is extremely safe: contributions to electro-weak observables
are suppressed and flavor physics is structurally protected. Yet such dynamics could very well lie at the TeV scale.

According to the standard lore a major drawback of this framework is the necessity of an elementary Higgs.
Nevertheless, while these theories have no direct motivation to explain the hierarchy between the Planck and electro-weak scale, they have appeared 
in various contexts, in particular to explain DM as accidentally stable composite state \cite{ACDM}, and for the relaxation of the electro-weak scale \cite{relaxion}. 
Moreover they could be remnants of SUSY or composite dynamics stabilising the electro-weak scale at a higher scale.

In this work we will explore the possibility of realizing composite Higgs models within this framework. While the subject of composite Higgs models
has strongly relied on effective theories with ad hoc properties, our framework provides a viable and calculable UV completion (see \cite{Marzocca:2013fza,vonGersdorff:2015fta,FPCstrumia} for a different  realisation involving partial compositeness). The lightest states are Nambu-Goldstone bosons (NGB) with quantum numbers determined by the charges of the vector-like fermions under the SM. A scalar NGB with quantum numbers of the Higgs doublet exists whenever Yukawa couplings between the fermions and an  elementary Higgs is allowed.

In isolation, fermion mass terms and SM gauge loops align the vacuum so that the electro-weak symmetry is not broken, i.e. the composite Higgs has no VEV.
The elementary Higgs allows a misalignment of the composite Higgs vacuum and the tuning of the electro-weak VEV. 
The system so constructed contains  an elementary Higgs doublet  and (at least) a composite one\footnote{This bears some resemblance with theories of induced electro-weak symmetry breaking where a sector that breaks electro-weak symmetry induces the expectation value for the elementary Higgs, see for example \cite{luty1,luty2}. In our construction  the strong dynamics does not break electro-weak symmetry so that it can be decoupled to arbitrarily high energies although at the price of tuning.}. 
Depending on the mixing induced by the Yukawa interactions, the lightest Higgs doublet can be mainly elementary or composite.
In \cite{Antipin:2015jia} we studied the possibility that the elementary Higgs has a negative mass parameter and induces electro-weak symmetry. 
We here turn to the regime where the Higgs VEV is induced through the mixing  with the (heavy) elementary Higgs. 
Curiously, this mechanism of electro-weak symmetry breaking was advocated in the first paper on composite Higgs models \cite{Kaplan:1983fs} and quickly forgotten.  One very attractive feature, though, is that the presence of an elementary Higgs allows to generate Yukawa couplings for the SM fermions in the standard way avoiding the severe problems with flavor physics of other composite Higgs models.

We will study under what conditions viable scenarios can be constructed. In particular, if the Higgs is mostly composite,  bounds from the $T$ parameter
imply that the strong dynamics should preserve custodial symmetry, a situation more easily realized in $\SO(N)$ and $\SP(N)$ theories. Requiring that the theory is valid up to a  high scale implies that the Higgs cannot be arbitrarily composite. The very same type of gauge theories discussed here have been used to realize the dynamical relaxation of the electro-weak scale \cite{relaxion}. We then explain how that mechanism can be realized in the present context giving an alternative explanation for  the electro-weak scale. 

The paper is organised as follows: In section \ref{sec:models} we review the general structure of the models discussing important constraints from the $T$ parameter, Higgs potential and perturbativity at high energy.  In section \ref{sec:LN} we consider in detail models based on $\SP(N)$ gauge theories with 4 fundamental fermions that produce a single Higgs doublet with custodial symmetry. We extend the analysis to $\SO(N)$ models in section 4. In section \ref{sec:relaxion} we explain how the relaxation of the electro-weak scale could work in this context, and we conclude in section \ref{sec:conclusions}. 
%\vspace{.2cm}

{\bf Note added:} During the final stages of preparation for this paper, Ref. \cite{Galloway:2016fuo} appeared with some overlap with section 3.

\section{Composite Higgs from confining gauge theories}
\label{sec:models}

We extend the SM with a new ``dark''  gauge group and fermions that are in a vectorial representations under the SM \cite{sundrum}.
We focus on the $\SU(N)$,  $\SO(N)$ and $\SP(N)$ gauge theories with fermions in the fundamental representation. 
The new dynamics is described by the renormalizable Lagrangian
\beq 
\Lag = \Lag_{\rm SM}    +  \bar\Q_i( i\slashed{D}  - m_i ) \Q_i - \frac{{{\cal G}_{\mu\nu}^{A2} }}{4g_D^2} 
+\frac {\theta_D} {32\pi^2} {\cal G}^A_{\mu\nu}\tilde{\cal G}^A_{\mu\nu}
+ [ H\bar\Q_i(y^L_{ij}P_L+ y^R_{ij}P_R) \Q_j + \hbox{h.c.}]
\label{lagrangian}
\eeq
The  Yukawa interactions with the Higgs doublet $H$ are possible depending on the quantum number of the fermions in the theory, while 
no Yukawa couplings between SM and vectorial fermions are allowed at the renormalizable level.
The  topological term for  dark gauge fields, which will only be relevant in the last part of this work, is physical for non-vanishing dark-quark masses $m_i$.

We assume that the new gauge dynamics is asymptotically free and confines at a scale $m_\rho$, spontaneously 
breaking the global symmetries as reviewed in  \cite{Witten:1983tx}.
We will be interested in the situation where the fermion masses are smaller than the confinement scale. 
Explicitly we will consider $\SU(N)$, $\SO(N)$ and $\SP(N)$ theories with fermions in the fundamental representation that
give rise to the following symmetry breaking patterns, 
\begin{equation}
\frac {\SU(N_F)\times \SU(N_F)}{\SU(N_F)}\,,~~~~~~\frac {\SU(N_F)}{\SO(N_F)} \,,~~~~~~\frac {\SU(N_F)}{\SP(N_F)} \,.
\end{equation}
All these cosets have the special property of being symmetric cosets. This simplifies the construction of the effective 
Lagrangian and we provide a unified description below. These theories are expected to follow the power counting of strongly coupled theories
\cite{SILH}. In particular we will assume the standard large $N$ scaling,
\begin{equation}
m_\rho = g_\rho f\,,~~~~~~~~~~~~~~~~~~~~~~g_\rho\sim \frac {4\pi}{\sqrt{N_D}}
\end{equation}
where $N_D$ is number of dark colors and $f$ the dynamical symmetry breaking scale.

We will further assume that the fermions belong to a vectorial representation of the SM. Since  gauge interaction and fermion 
masses align the vacuum of the theory in the unbroken direction, this guarantees that the strong dynamics does not  break the SM electro-weak symmetry in isolation.
For this reason it is necessary to include an elementary Higgs to trigger the electro-weak symmetry breaking. 

\subsection{Yukawa Couplings}

Of particular relevance for the present work will be the couplings of the elementary Higgs with the vector-like fermions.
Whenever the quantum numbers allow for Yukawa couplings, a NGB with the same quantum number as the Higgs exists in the spectrum. We will call it $K$. 
This simple group theory fact, that follows immediately from the fact that NGBs have quantum numbers of pairs of fermions, implies that
one effectively obtains a  2 Higgs Doublet Model (2HDM) of type I, given that the SM fermions can only couple to the elementary Higgs at renormalizable level. Turning this around, any gauge theory that delivers a Higgs as a NGB allows a Yukawa coupling with an elementary Higgs.

Elementary and composite Higgs will mix as dictated by the Yukawa couplings so that the physical Higgs is a mixture of $H$ and $K$.
The same setup was studied in \cite{Antipin:2015jia, Antipin:2014qva} in the limit where the 125 GeV Higgs is mostly elementary. 
We here turn to the composite regime where the elementary Higgs is necessary to induce the electro-weak symmetry breaking for $K$. 
Moreover, the presence of $H$ allows to generate SM Yukawas coupling, avoiding all flavor problems.

Schematically, the physics of these models can be understood as follows:
similarly to QCD, Yukawa coupling of the underlying theory $y H\bar\Q \Q$ matches to the 
effective coupling $y m_\rho f H K$ where we used the fact that the fermion bilinear interpolates with the NGB as $K\sim \bar\Psi \Psi/{m_{\rho}f}$.
Then, the mass matrix for elementary and composite Higgses has the form,
\begin{equation}
M^2 = \left( \begin{array}{cc}
m_H^2 & \epsilon m_H^2 \\
\epsilon^* m_H^2 & m_K^2  \end{array} \right) .
\label{mass22}
\end{equation}
where the mixing parameter $\epsilon$ has the structure,
\begin{equation}
\epsilon\sim (y-\tilde{y}^*) \frac {m_\rho f}{m_H^2},
\end{equation}
and where $y$ and $\tilde{y}$ are the Yukawa couplings of the two chiralities of the vector like fermions.
For $y=\tilde{y}^*$ the couplings preserve parity. In this case, the mixing vanishes because the Higgs is scalar and $K$ is pseudo-scalar, 
effectively realizing an inert 2 Higgs doublet model.
%also note that one may choose to factor out $m_K$ instead of $m_H$ to form the dimensionless mixing parameter $\epsilon$, depending on what hierarchy between the two mass parameters one is interested in, as in the previous study \cite{Antipin:2015jia}.
The composite Higgs acquires calculable contributions to its mass from all the effects that explicitly break the global symmetries,
i.e. gauge and Yukawa couplings and fermion masses. One finds,
\begin{equation}
m_K^2 \sim \frac {g^2}{16\pi^2} m_\rho^2 +\frac  {y^2}{16\pi^2} m_\rho^2 +  m m_\rho ,
\end{equation}
where we wrote collectively the dark-quarks masses $m_i$ as $m$.

In order to break the electro-weak symmetry and obtain $v\ll f$ one should require Det$[M^2]\approx 0$ and negative, i.e.
\begin{equation}
m_K^2  \approx |\epsilon|^2  m_H^2
\label{eq:tuning}
\end{equation}
This condition, that needs to be satisfied with a precision $v^2/f^2$, is the tuning of the electro-weak VEV in composite Higgs models. 
Using the equation above, the mixing angle between elementary and composite Higgs reads
\begin{equation}
\tan \beta \approx |\epsilon|\approx \frac {m_K^2}{|y-\tilde{y}^*| m_\rho f}.
\end{equation}
Given that the gauge contribution implies $m_K> m_\rho/10$, it follows that the physical Higgs can be mostly composite for $y, \tilde{y}\sim 1$ or even smaller.

When $|\epsilon| < 1$, the physical Higgs will be mostly composite. The couplings of the elementary Higgs with SM fermions
generates at low energies the Yukawa couplings for the composite Higgs,
\begin{equation}
y^{SM}\approx y^{EL} \sin \beta .
\label{SMyukawa}
\end{equation}
This implies that, in order to reproduce the known SM Yukawa couplings, the elementary ones must be larger according
to the mixing. For this reason the mixing cannot be arbitrarily large in order to avoid perturbativity bounds, in particular from the top quark. 
Note however that the flavor structure is identical to the SM so that all flavor violation is controlled by the SM Yukawas. 
This allows to easily avoid all flavor bounds that plague ordinary composite Higgs models.

Discrete symmetries play an important role in this framework. For $y\ne \tilde{y}^*$ the theory violates parity.
Since the elementary Higgs is even under parity while NGBs are odd, their mixing is proportional to the violation of 
parity. Moreover for complex Yukawas an extra CP violating phase exists in the theory. This has important physical effects, generating in particular an Electric Dipole Moment (EDM) for the electron, that could be potentially observed.

\subsection{T-parameter}

In general, the strong dynamics that produced the Higgs as NGB might not respect custodial symmetry. 
Higher dimensional operators produce tree-level corrections to the $T$-parameter that scale as
\begin{equation}
\Delta T \sim \frac {v^2}{f^2} .
\end{equation}
At face value, this implies  $f>5$ TeV to comply with experimental constraints.

In order to construct a viable composite Higgs model with relatively low symmetry breaking scale, say $f\sim 1$ TeV, the Higgs
should be  in the $(2,2)$ representation of custodial symmetry $\SU(2)_L\times \SU(2)_R$ \cite{SILH}. This  means that
the unbroken group should contain the full custodial symmetry, a condition  possible only if the constituents are complete representations 
of custodial symmetry.

In $\SU(N)$ gauge theories with fermions in the fundamental, the NGBs have the quantum numbers of the product of left-handed and right-handed fields.
The simplest way to obtain a bi-doublet is to consider doublet and singlet of $\SU(2)$. There are two possibilities:
\begin{equation}
\Psi=(2,2) + (1,1)\,,~~~~~~~~~~~~~~~~~\Psi=(2,1) + (1,2) ,
\label{custodialrep}
\end{equation} 
where each fermion is Dirac. In both cases one obtains 2 Higgs doublets in (2,2) rep. The only way to obtain a single 
Higgs doublet is when it transforms as  $(2,1)_{\frac 1 2}$,  which is not a custodial preserving rep. This is for example the case of the 
coset $\SU(3)_L\times \SU(3)_R/\SU(3)$ discussed in \cite{Antipin:2015jia}. This feature is associated to the fact that the fundamental rep of $\SU(N)$  is complex so that  the $(2,2)$ NGB are necessarily in a complex representation.

In general, with more than one Higgs doublet custodial symmetry $\SU(2)_L\times \SU(2)_R$ is not sufficient \cite{Mrazek:2011iu}. The reason is the following: 
after electroweak symmetry breaking the $W$ gauge bosons should transform as a triplet of $\SU(2)_c$ so that $m_W^2/m_Z^2 \cos^2 \theta_W\approx 1$. 
$\SU(2)\times \SU(2)$ is isomorphic to $\SO(4)$ which is the symmetry group of the 3-sphere $S_3$. 
A Higgs VEV picks a direction on the sphere breaking $\SO(4)$ to $\SO(3)$.
A second Higgs acquiring a generic VEV would than break $\SO(3)$ to $\SO(2)$ leading to tree level corrections to the $T$ parameter (assuming it 
doesn't break charge).  In the models above we can write custodially preserving Higgs couplings,
\begin{equation}
(2,2) H (1,1)\,,~~~~~~~~~~~{\rm or}~~~~~~~~~~~ (2,1) H (1,2)
\end{equation}
that would lead to a custodial preserving vacuum. Note however that compatibly with SM symmetries  
one could write different Yukawa couplings for the two chiralities of the vector like fermions that explicitly break custodial symmetry. 
The potential generated would be such that the minimum does not preserve custodial symmetry.
In this case corrections to $T$ of order $v^2/f^2$ will be generated.

The situation is different in $\SO(N)$ and $\SP(N)$ models as in this case the representations are real (or pseudo-real).
In $\SO(N)$ theories we can take the first rep in (\ref{custodialrep}) to build a real representation. This will produce a single 
Higgs doublet in (2,2) rep of custodial symmetry. The second representation can instead be used in $\SP(N)$ theories. The reason for this
is that combining the reps of $\SP(N)$ with the one of $\SU(2)$ one can build a real representation\footnote{The second rep would not work
for $\SO(N)$ and the first for $\SP(N)$ because they would not produce a real representation.}. Even with these protected representations
the Yukawa couplings can break custodial symmetry leading to important tree level effects.

While the $T$-parameter can be protected through custodial symmetry, other contributions to the precision tests are the same as 
in composite Higgs models. In particular the contribution to the $S$ parameter from vector resonances $\Delta S\sim 4\pi v^2/m_\rho^2$
implies that the  dynamical scale should be in the multi TeV range.

\subsection{Higgs Potential}
\label{sec:higgs}

A notable feature of the NGB Higgs is that, once the electro-weak VEV is tuned, the Higgs mass is predicted in terms of the couplings that break explicitly the global symmetries.  Having an underlying renormalizable theory allows then in principle to derive the Higgs mass from the fundamental parameters of the theory. 
We here sketch the general structure of the potential before discussing explicit examples.

In the models under consideration there are in general 3 contributions to the composite Higgs potential: gauge loops, 
vectorial fermion masses and  Yukawa couplings with the elementary Higgs. The first two in general align the vacuum in the unbroken direction
while the latter tends to destabilize it. As a consequence, by tuning coefficients, one can make $v/f \ll 1$ as required for the composite Higgs.

First we notice that for a symmetric coset $G/H$ such as the ones we are considering the NGB can be described by the matrix,
\begin{equation}
\Sigma = U(\Pi)^2\,,~~~~~~~~~~~~~~~~U(\Pi)=e^{i \Pi}
\end{equation}
where $\Pi$ contains the broken  generators. The advantage of using $\Sigma$ is that it transforms linearly under the full $G$ group 
making it easier to write invariants. The potential has the following structure,
\begin{equation}
V\sim  A\, {\rm Tr}[M(H)\cdot \Sigma\cdot \Sigma_0+ h.c.] +B\,\sum_a  g_a^2\,{\rm Tr}  [T^a \Sigma {T^a}\Sigma^\dagger]
\end{equation}
where,
\begin{equation}
M(H)= (m+ h^a Y^a )\,,~~~~~~~~~~~~~~~~~~(\Sigma_0)^{ij}= \langle \Psi^i \Psi^j\rangle
\end{equation}
and $g_a$ are the SM couplings.  Here $m$ is the diagonal matrix of vectorial masses and the matrices $Y^a$ contain the Yukawa couplings to the corresponding component $h^a$ of $H$. The naive dimensional analysis estimate of the coefficients is
$A=m_\rho f^2$ and $B=m_\rho^2 f^2/(4\pi)^2$.

The potential has a special structure compared to a generic 2HDM since the elementary Higgs only appears linearly in the leading order terms. As a consequence quartic couplings such us $H^2 K^2$ or $H^3 K$ are suppressed compared to $H K^3$. Integrating out at tree level the elementary Higgs doublet we obtain\footnote{We have omitted from the effective potential contributions from loops with Yukawa couplings.
These can  be computed in terms of form factors of the strong dynamics as explained in \cite{Antipin:2014qva}. A spurionic analysis shows that the functional structure is identical to the one induced integrating out the elementary Higgs.},
\begin{equation}
V(\Pi)\approx A {\rm Tr}[m_0\cdot \Sigma\cdot \Sigma_0+h.c. ]- \frac {A^2}{2m_H^2} \sum_a|{\rm Tr}[Y^a\cdot\Sigma\cdot \Sigma_0]|^2 + B\,\sum_a  g_a^2\,{\rm Tr}  [T^a \Sigma {T^a} \Sigma^\dagger]
\end{equation}
Since the potential contains terms either linear or quadratic in $\Sigma$, projecting onto the physical Higgs direction one obtains the following terms:
\begin{equation}
V(h) \approx  a \cos \frac h f + b \sin^2 \frac  h f
\end{equation}
where $a$ and $b$ are linear combinations of the coefficients in $V(\Pi)$.
Demanding that the electroweak VEV is reproduced, allows to determine the Higgs mass as,
\begin{equation}
\cos \frac {\langle h \rangle} f= \frac {a} {2 b}\,,~~~~~~~~~~~~~ m_h^2= - \frac {2 b}{f^2}\sin^2 \frac {\langle h \rangle} f 
\end{equation}
The first condition  is the tuning of order $v^2/f^2$ on the coefficients $a$ and $b$ that needs to be performed to obtain a small Higgs VEV. 
The second is the prediction of the Higgs mass in terms of the coefficients of the effective Lagrangian that are fully determined by the strong dynamics.

In addition if the elementary Higgs has a quartic term this produces an extra contribution to the Higgs mass,
\begin{equation}
\Delta m_h^2  \approx 2 \lambda v^2 \sin^4 \beta \approx 2\lambda v^2 |\epsilon|^4
\end{equation}
Being suppressed by 4 powers of the mixing this is typically negligible. The presence of a quartic coupling is however important for the high energy behaviour of the theory.

\subsection{Running}
\label{sec:running}

So far our discussion was at the level of the effective theory. 
To establish the validity of the UV completion we now study the evolution of the couplings of the microscopic Lagrangian at high energies. 
%We parametrize the Yukawa couplings as in \eqref{Sp4Yukawa} but our results apply in general for theories described by \eqref{lagrangian} with minor modifications.  
At energies higher than the confinement scale the evolution of the couplings is determined by the $\beta-$functions,
\begin{eqnarray}
 \frac{d y_t}{d\log\mu} &=& \frac{y_t}{(4\pi)^2} \bigg[ \frac 9 2 y_t^2 + N_D( |y|^2 +|\yt|^2) - 8 g_3 ^2- \frac{17}{12} {g'}^2-\frac{9}{4} g^2 \bigg]  \nonumber \\
\frac{d y}{d\log\mu}&=& \frac {y}{(4\pi)^2} \bigg[ \frac 3 2 (|y|^2-|\yt|^2) + N_D( |y|^2 +|\yt|^2) +3 y_t^2 - 6\frac{\mathrm{dim}(G_D) T(R)}{N_D} g_D ^2  \bigg]  \quad \beta_{\yt} = \beta _y (y\leftrightarrow \yt) \nonumber \\
\frac{d\lambda}{d\log\mu}& =& \frac {4}{(4\pi)^2 }\bigg[3\lambda^2 - (3 y_t^4+ N_D (|y|^2 +|\yt|^2)^2)+\lambda (3 y_t^2+ N_D (|y|^2 +|\yt|^2)) \bigg] \nonumber \\
\frac{d g_D}{d\log\mu}&=&- \frac {g_D^3}{(4\pi)^2} \bigg[ \frac{11}{3} C_2(G)-\frac23 T(R) N_f \bigg]
 \end{eqnarray}
where $N_f$ is the number of Weyl fermions and,
\beq
T(R)= \{1/2, 1, 1/2\} \qquad \qquad C_2(G)=\{N_D,N_D-2, N_D+1\}
\eeq 
for $\SU(N)$, $\SO(N)$ and $\SP(2N)$ respectively.
The main difference from the SM originates from the boundary condition in \eqref{SMyukawa}. That equation should be understood at the confinement scale.
Because of the enhanced value of the Yukawa coupling compared to the SM, the  top Yukawa may run into Landau poles at relatively low scales if the
Higgs is mostly composite. At face value for $y, \tilde{y}\ll 1$ (also favoured by Higgs mass considerations), 
a negative $\beta_{y_t}$ function at a scale $\sim \mathcal{O}$(few TeV) would require,
\beq
\sin^2\beta \gtrsim \frac{9}{2\big( 8 g_3^2+ \frac{17}{12} {g'}^2+\frac{9}{4} g^2\big)}\approx 0.5  \qquad (i.e.\  \tan\beta \gtrsim 1)\ .
\label{two-loop}
\eeq
%However, since the $g_{\rm QCD}$, which dominates numerically, decreases as we increase the energy scale, the actual value of the $\tan\beta$ needed for the theory to be valid up to the Planck scale is bigger. 
\begin{figure}[h!]
\begin{center}
\includegraphics[width=0.68\textwidth]{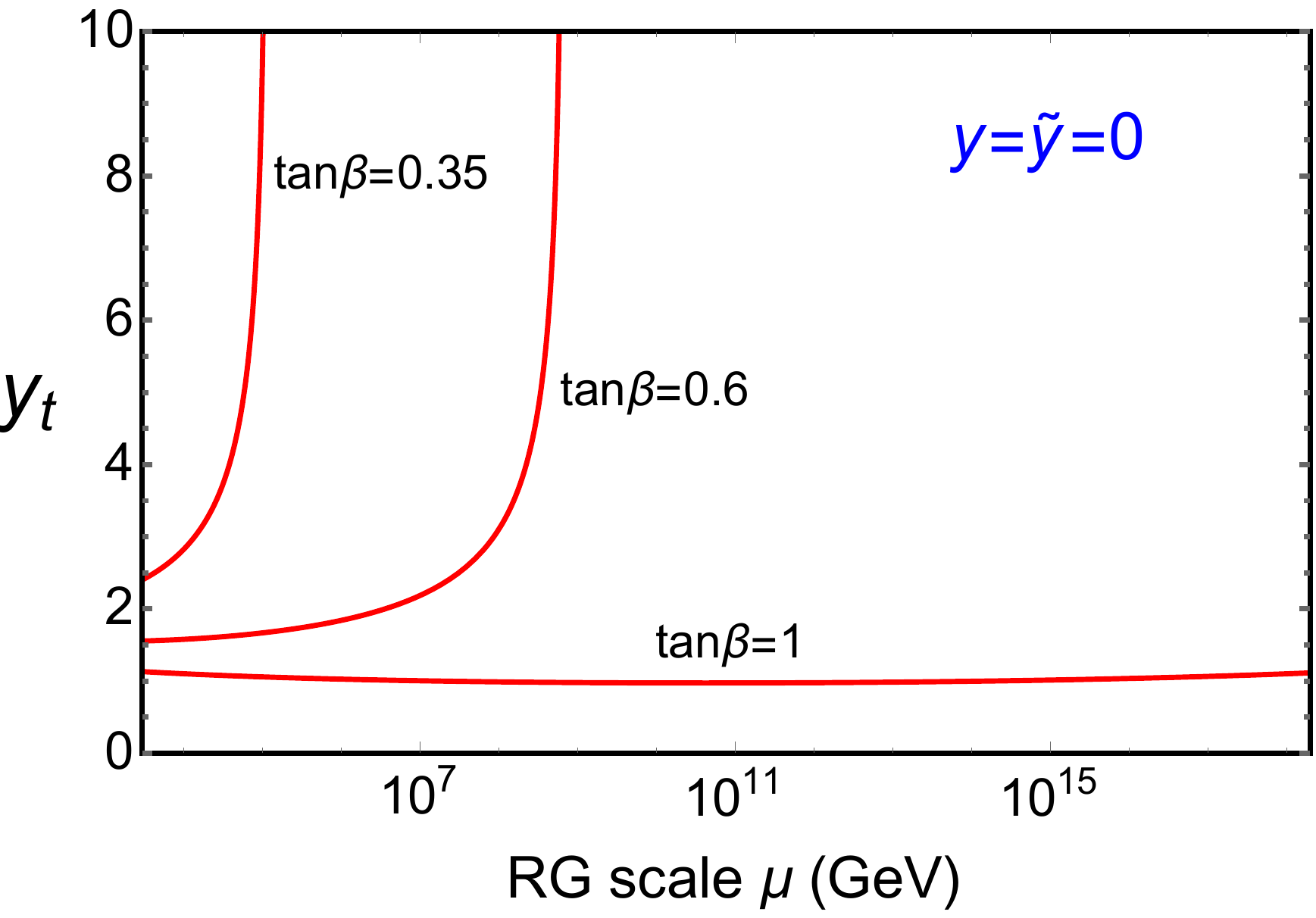}
\caption{Evolution of the top Yukawa couplings at high energy (1-loop) for $y=\tilde{y}=0$ with $N_D=2$. We assumed the 
boundary condition for the top Yukawa $y_t$(3 TeV)$\approx 0.8 /\sin\beta$.}
\label{fig:running}
\end{center}
\end{figure}
In Figure \ref{fig:running} we consider the evolution of top Yukawa coupling as a function of the mixing angle between elementary and composite Higgs. For $\tan\beta=1$ and $y=\tilde{y}=0$ the theory could be valid up to the Planck scale while a lower scale is obtained when the Higgs is more composite. We have checked that this result remains valid as long as the dark Yukawas $(y,\tilde{y})\lesssim$ 0.5, which is satisfied in the concrete model examples later. 

Things could improve if also QCD becomes not asymptotically free 
due to extra colored matter so that $g_3$ increases at high energy.  Also, starting at two-loop order, there is a $\sim N_{D} g_D^2( |y|^2 +|\yt|^2)$ contribution from the dark gauge coupling to $\beta_{y_t}$ that may change the estimate in eq. (\ref{two-loop}) since at the confinement scale $g_D$ is large. We leave a detailed study of these possibilities to future work.

Concerning the quartic of the elementary Higgs, the top quark and the new SM fermions give a large negative contribution to
the $\beta$ function of $\lambda$. Therefore if $\lambda(m_\rho)=0$ the quartic becomes quickly unstable developing a new minimum.
This can be easily remedied introducing a tree level quartic. In the composite regime, given that the contribution to the Higgs squared mass is suppressed by four 
powers of the mixing, this affects little the prediction of the Higgs mass. For example for $\tan\beta=1$ and $(y, \tilde{y}) \ll 1$ then $\lambda(3 \ {\rm TeV})\approx 0.8$ in order to  avoid instabilities or Landau poles. Nevertheless issues about stability of the SM vacuum need to be reconsidered in this context.

\section{$\SU(4)/\SP(4)$ model}
\label{sec:LN}

In this section we will consider the most economical gauge  theory that produces a composite Higgs doublet with custodial symmetry.
This has been considered in the past by several groups \cite{Gripaios:2009pe,CTC,Cacciapaglia:2014uja,Ferretti:2016upr,Alanne:2014kea,Gertov:2015xma}.
The pattern of symmetry breaking $\SU(4)/\SP(4)$  can be realised by an $\SP(N_D)$ dark gauge group with 2 Dirac fermions in the
$N_D$ dimensional rep of the group.  This rep is pseudo-real so that the fermion kinetic terms enjoy an enhanced flavor symmetry $\SU(4)$ compared to the apparent $U(2)_L\times U(2)_R$ in terms of Dirac fermions. 
The gauge dynamics generates a fermion condensate that breaks spontaneously the symmetry to $\SP(4)\simeq \SO(5)$ delivering 5 NGB.

Let us denote the 2 Dirac flavors as $U$ and $D$. In terms of left-handed Weyl spinors these are
\beq
\Psi \equiv
\begin{pmatrix}
U_L\  D_L \ U_R \ D_R
\end{pmatrix}^\intercal .
\eeq
To embed the electro-weak symmetry inside the global $\SU(4)$ symmetry  $Q_L\equiv (U_L\ D_L)^\intercal$ transforms as a doublet of SU(2)$_L$ and $Q_R\equiv (U_R \ D_R)^\intercal$ as a doublet of SU(2)$_R$ :
%(the prime denotes the conjugate field, with both Lorentz and color anti-symmetrized such that it transforms the same as $Q_L$ under Lorentz and dark-color transformations):
\beq
Q_L = ( {\bf 2}, {\bf 1} )\,, \qquad Q_R = ({\bf 1}, {\bf 2})\,,
\eeq
%where we now have 4 left-handed Weyl fermions, and the quantum numbers refer to SU(2)$_L \times$ SU(2)$_R$ with 
the hypercharge being the diagonal generator of SU(2)$_R$.
%This assignment is chosen so that we keep track of the custodial symmetry and have the PNGB containing the object $Q_L Q_R$, which transforms like the SM Higgs field. 
Quantum numbers allow masses $\mu_L$ and $\mu_R$ for the two doublets.  In the limit $\mu_L = \mu_R$, the quark masses are  aligned with our starting vacuum and do not break the global symmetry further. For generic complex masses $\mu_L$ and $\mu_R$, instead, the residual symmetry is SU(2)$_L \times$ SU(2)$_R$. Moreover independent Yukawa couplings with the elementary Higgs can be written down for $Q_L U_R$ and $Q_L D_R$.
  
%In summary, besides dark gauge interactions and the usual Standard Model Lagrangian with elementary Higgs doublet $H$, the only interactions allowed by all the symmetries are Yukawa interactions of the new fermions.
%Combining the four Weyl fermions as
%\beq
%\Psi \equiv
%\begin{pmatrix}
%U_L\  D_L \ U_R^\prime \ D_R^\prime
%\end{pmatrix}^\intercal, \quad
%\text{and} \quad
%\Psi^\prime \equiv
%\begin{pmatrix}
%U_L^\prime\  D_L^\prime \ U_R \ D_R
%\end{pmatrix}^\intercal,
%\eeq
The mass terms and Yukawa interactions take the form
\beq
\mathcal{L}_{\rm} = - \frac{1}{2}\, {\Psi}^\intercal M \Psi + {\rm h.c.}
\label{Yukawatree}
\eeq
with the anti-symmetric matrix
\beq
M =
\begin{pmatrix}
0 & \mu_L &
\multirow{2}{*}{{$-\yt \tilde{H}^*$}} & \multirow{2}{*}{{$y H^*$}} \\
-\mu_L & 0
& & \\
\multicolumn{2}{c}{\yt \tilde{H}^\dagger}
& 0 & -\mu_R \\
\multicolumn{2}{c}{-y H^\dagger}
& \mu_R & 0
\end{pmatrix}
\label{Sp4Yukawa}
\eeq
where $H$ is the Higgs doublet and $\tilde{H}^i=\epsilon^{ij} H^j$. Note  that custodial invariance  requires $\yt = -y$.

For generic dark quark masses, we can use two phases of the 3 fermion fields (doublet $Q_L$ and singlets $U_R$, $D_R$) to make the mass parameters real. 
The remaining fermion phase can be used, for instance, to make $y$ (or $\yt$)  real, while the relative phase between $y$ and $\yt$ is physical and could in principle lead to CP-violating effects.
We will however see that such phase does not appear in the lowest order chiral Lagrangian.

\subsection{Effective Lagrangian}

Following section \ref{sec:higgs} we can easily write the effective Lagrangian for the composite Higgs.
As reference vacuum we can choose \footnote {The most general choice (up to a phase) of the vacuum breaking $\SU(2)_L$ can be parametrized as
 \beq
\Sigma_0 = \left( \begin{array}{cc}
i \sigma_2  \cos\theta& \sin\theta \\
-\sin\theta & - i \sigma_2 \cos\theta
\end{array} \right)
\eeq
where the angle $\theta$ describes how the condensate aligns itself in the direction that does not break the gauge symmetry ($\theta = 0$) and the `technicolor' limit ($\theta = \pi/2$) passing through the composite Higgs limit $\theta\ll 1$. In this work we will concentrate on the composite Higgs limit and this is why we choose the $\theta=0$ vacuum and parametrise the misalignment as a vacuum expectation value of the pion field rather than a change in $\Sigma_0$. We will comment on the `technicolor' limit below.}
\beq
\Sigma_0 = \left( \begin{array}{cc}
i \sigma_2  & 0 \\
0 & - i \sigma_2 
\end{array} \right)\,.
\label{SP4vacuum}
\eeq
We then parametrize the 5 NGB as
\beq
\Sigma \equiv
\exp\left[\frac{2\sqrt{2}\,i }{f} \sum_{a=1} ^{5} X^a \Pi_a\right] \cdot \Sigma_0
\eeq
where the generators of $\SU(4)$ are normalized as Tr($T^i T^j$) = 1/2 $\delta^{ij}$;
we denote the generators broken by the vacuum $\Sigma_0$ as $X^i$ and the unbroken ones as $S_i$, using the same basis as \cite{Cacciapaglia:2014uja}.

The two derivative low energy Lagrangian, together with the elementary Higgs one, reads
\bea
\label{lagr}
\mathcal{L}&=&\ \frac{f^2}{8}\, \mathrm{Tr}\left( \left(D_{\mu} \Sigma\right)^\dagger D^{\mu} \Sigma \right) - \frac{1}{2} C_y f^3 \big(\mathrm{Tr}[M \Sigma] + h.c.\big) - C_g f^4 g_i^2 \mbox{Tr} [S^i\, \Sigma\, {S^i}^*\, \Sigma^\dagger]
\nonumber \\
&+&\ \left(D_\mu H\right)^\dagger D^\mu H - \mu_H^2 H^\dagger H - \lambda \left( H^\dagger H \right)^2 
\eea
where we expect the scaling $C_y \sim m_\rho/f$, $C_g \sim m_\rho^2 / (4\pi f)^2 \sim 1/N_D$ \cite{Antipin:2014qva}.
The covariant derivatives are defined as\footnote{We denote $A_\mu=  g \,W _\mu ^i S^i _L + g' B_\mu S^6 _ R$ and $\mathcal{A}_\mu = g \, \frac{\sigma_i}{2}W_\mu ^i + \frac{1}{2}\,g' \mathbb{1}_2 B_\mu$ depending on whether they act on 4 dimensional or 2 dimensional vectors.}
\begin{align}\label{su4:cov}
D_{\mu} \Sigma&=\partial _\mu \Sigma -  i  \left( A_\mu  \Sigma + \Sigma A_{\mu}^\intercal \right), \\
D_\mu H &= \partial _\mu H -i\, \mathcal{ A}_\mu  \, H\, .	\nonumber
\end{align}

The terms proportional to $C_y$ and $C_g$ above are the contribution of the masses and gauge loops respectively (with $S^i$ being the corresponding generators with gauge coupling $g_i$).
%(in the notation of \cite{Cacciapaglia:2014uja}, the gauge generators of $\SU(2)_L$ are $S^{1,2,3}$ while the one for $U(1)_Y$ is $S^6$) 
%while the second term is the Wess-Zumino-Witten anomaly of the NGB $\eta\equiv \Pi_5$ (which couples to the generator $Y^5$) with two SM gauge bosons. 
%Explicitly, the anomalous couplings  are (see \eqref{vewdef} below for the definition of the angle $\theta$) \cite{Ferretti:2016upr}
%    \begin{gather}
%     g_{\eta WW}=\frac{N_D g^{2}\cos{\theta}}{16 \pi^{2}f}\,,
%     \qquad
%     g_{\eta ZZ}=\frac{N_D (g^{2}-g'^{2})\cos{\theta}}{32\pi^{2}f}\,,
%    \qquad
%     g_{\eta Z\gamma}=\frac{N_D gg' \cos{\theta}}{16 \pi^{2}f}\,, \nonumber \\
%     g_{\eta\gamma\gamma}=0\,,
%     \qquad
%     g_{\eta g g}=0\,.
%    \end{gather}
%The coupling to gluons vanishes as the dark quarks do not carry QCD colour, while the coupling to photons vanishes due to the fact that U(1)$_{em}$ is fully embedded in SU(4).
%These $\eta \to g g$ and $\eta \to \gamma \gamma$ couplings will be
%generated by the top quark loop via the $\eta\bar{t}t$ coupling which is proportional to the Sp(4)-violating combination of the dark quark masses $\mu_{L,R}$
% at the next-to-leading order 
%\cite{Arbey:2015exa}. This combination is expected to be small (compared to the Sp(4)-preserving combination) as otherwise the symmetry
%breaking pattern SU(4)/Sp(4), that gives rise to a NGB Higgs, would be distorted.

The 5 NGB decompose under the electro-weak symmetry as a complex doublet $K$ and a singlet $ \eta $.
The Lagrangian in eq. (\ref{lagr}) can be written explicitly as
\bea
\label{lagrexpand}
\mathcal{L}&=&
\, \frac{\sin^2{S}}{S^2} \left( \left(D_\mu K\right)^\dagger D^\mu K + \frac{1}{2}\left( \partial_\mu \eta \right) ^2\right) \ +\frac{S^2-\sin^2 S}{2 f^2 S^4} \left(K^\dagger \partial_\mu K + \left(\partial_\mu K\right)^\dagger K+\eta\cdot \partial_\mu \eta \right)^2   \nonumber
\\
&&+
\left( \mu_H^2 \left\{ \epsilon\, H^\dagger K + \mathrm{h.c.} \right\}
- 2 C_y\, \mathrm{Im}(\mu_L + \mu_R^*) f^2\, \eta \right)
\cdot \frac{\sin{S}}{S}
\\
&&+\ \mu_\eta^2\, f^2 \cdot \cos{S}
+\ \frac{1}{2} \mu_g^2 \left( f^2 \cos^2{S} + \eta^2\, \frac{\sin^2{S}}{S^2} \right)
\nonumber\\
&&+\ \left(D_\mu H\right)^\dagger D^\mu H - \mu_H^2 H^\dagger H - \lambda \left( H^\dagger H \right)^2 \nonumber
\eea
where we introduced $S\equiv \frac{\sqrt{2 K^\dagger K + \eta^2}}{f}$
and defined
\beq \label{eq:su4potdef}
\epsilon \equiv -\sqrt{2} i(y-\yt^*)\, \frac{C_y f^2}{\mu_H^2},\
\mu_\eta^2 \equiv 2 C_y \mathrm{Re}(\mu_L + \mu_R^*) f,\
%\text{ and }
\mu_g^2 \equiv C_g (3 g^2+{g'}^2) f^2,
%\text{ and for later }
%\mu_K^2 \equiv \mu_\eta^2 + \mu_g^2.
\eeq
%First two lines are obtained by expanding the kinetic term for $\Sigma$ and we notice that pions kinetic terms are not canonically normalized and need to be redefined in order to absorb the factor $\sin{S}/S $. 
and, for convenience, we will also use $\mu_K^2 \equiv \mu_\eta^2 + \mu_g^2$.

Electro-weak symmetry is broken by the vacuum expectation values $v_H/\sqrt2$ and $v_K/\sqrt2$ of  $H$ and $K$. One finds,
\beq\label{vewdef}
v_{EW}^2 = v_H^2 + f^2\,\sin^2{\theta}
\quad
\text{with}
\quad
\theta \equiv \frac{v_K}{f}.
\eeq
We also define $\tan{\tilde\beta}$ as the ratio of the contributions from both sectors to $v_{EW}$:
\beq
\tan{\tilde\beta} \equiv \frac{v_H}{f\sin{\theta}},
\label{eq:tanbtdef}
\eeq
such that $v_H = v_{EW} \sin{\tilde\beta}$ and $f \sin\theta = v_{EW} \cos{\tilde\beta}$.
We think of $\tilde\beta$ as an input parameter, but still use $\theta$ in expressions whenever convenient.

There is a linear coupling for $\eta$ proportional to  $\mathrm{Im}(\mu_L + \mu_R^*)$  which we remove by re-phasing the underlying dark quark fields, 
making the quark masses real, in order to have $\left<\eta\right>=0$.
Similarly, the combination $i(y-\yt^*)$ appearing in $\epsilon$ can be made real.
There are thus no more complex phases remaining in the Lagrangian.
As a consequence the effective Lagrangian respects CP despite the fact that the microscopic action contains a physical phase.
CP invariance is an accidental feature of the 2-derivative effective Lagrangian that does not persist to higher orders, see appendix.

The pion potential is readily obtained from the Lagrangian \eqref{lagrexpand}.
%\bea
%V
%&\approx&  \mu_H^2 H^\dagger H + \lambda \left( H^\dagger H \right)^2
%+ \mu_K^2 K^\dagger K + \mu_\eta^2 \frac{\eta^2}{2} \\
%&&-\ \epsilon\,\mu_H^2 \left( H^\dagger K + K^\dagger H \right) \Bigl( 1 - \frac{1}{3\,f^2} \left( K^\dagger K + \frac{\eta^2}{2} \right) \Bigr)
%\nonumber \\
%&&-\ \frac{1}{3\,f^2} \Bigl(
%\left( \frac{1}{2} \mu_\eta^2 + 2 \mu_g^2 \right) \left( K^\dagger K \right)^2
%+ \left( \mu_\eta^2 + 2 \mu_g^2 \right) K^\dagger K \frac{\eta^2}{2}
%+ \frac{\mu_\eta^2}{2} \frac{\eta^4}{4}
%\Bigr),
%\nonumber
%\eea
We want now to relate the parameters of the model to the vacuum expectation values.
The extrema of the potential are given by
\bea\label{eq:vevsp4}
 \mu_H^2\, v_H + \lambda\, v_H^3 &=& \epsilon\,\mu_H^2\, f \sin{\theta} \\
\label{eq:vevsp4:2}
\epsilon\,\mu_H^2\, v_H\, \cos{\theta} &=& \left( \mu_\eta^2 + \mu_g^2\, \cos{\theta} \right) f \sin{\theta},
\eea
which we can rewrite as
\bea
\label{eq:mHmK}
\mu_H^2 \tan^2{\tilde\beta} &=& \mu_K^2 + \mu_\eta^2 \left( 1/\cos\theta-1 \right) - \lambda\, v_{EW}^2 \sin^2{\tilde\beta} \tan^2{\tilde\beta}
\ \approx\ \mu_K^2 \\
\epsilon &=& \tan{\tilde\beta} \left( 1 + \lambda \frac{v_{EW}^2}{\mu_H^2} \sin^2{\tilde\beta} \right)
\ \approx\ \tan{\tilde\beta}.
\eea
The tuning $v_{EW} \ll f$ (i.e.~$\theta \ll 1$), in the regime where the $\lambda$ contribution is negligible, can be seen from equation \eqref{eq:mHmK};
the difference between $\mu_H^2 \tan^2\tilde\beta$ and $\mu_K^2$ is indeed proportional to the quantity we want to make small, $v_{EW}^2/f^2$, since from the definitions of $\theta$ and $\tilde\beta$ above
\beq\label{su4:tuning}
1/\cos\theta - 1 = \cos^2{\tilde\beta} \frac{v_{EW}^2}{2 f^2} + \mathcal{O}\left[ \frac{v_{EW}^4}{f^4}\right].
\eeq

Finally, also from \eqref{eq:mHmK}, $\mu_H^2$ does go to the expected Standard Model value $-\lambda v_{EW}^2$ in the `elementary' limit, $\tan\tilde\beta\to\infty$, however the point at which it turns negative corresponds to values of $\tilde\beta$ ever closer to $\pi/2$ as $f$ (thus $\mu_K$) is increased.

Finally let us comment on the value of $y$ determined by the choice of parameters (in particular of $\tilde\beta$).
Inverting the definition of $\epsilon$ \eqref{eq:su4potdef} and imposing the VEV equations just obtained, we find that
\beq
\mathrm{Im}[y-\yt^*] \sim \left( \text{gauge contribution} + \frac{\mu_L+\mu_R}{f} \right) / \tan\tilde\beta,
\eeq
where the gauge contribution coefficient is a few percent, and the quark masses, as we find in the next section, need to be fixed to few percent of $f$ as well, to reproduce the correct Higgs mass in the small-$\tilde\beta$ regime. Thus, as anticipated in Section~\ref{sec:running}, for moderately small values of $\tan\tilde\beta$, the Yukawa couplings are indeed less than one.

%As we are interested in the regime where $\tan{\tilde\beta}$ is as small as possible, and although we will see that it cannot in fact be made too small, it is instructive to integrate out the elementary field $H$, whose mass term grows as $1/\tan\tilde\beta$.
%\xxx{reference to bound from running. how do we link the sections?}
%The equations of motion give
%\beq
%H = \epsilon\, K\, \frac{\sin{S}}{S} + \mathcal{O}\left[\lambda\right]
%\approx \tan{\tilde\beta}\ K.
%\eeq
%In terms of the canonically normalised field $\tilde{K} \equiv K / \cos{\tilde\beta}$, the potential to quartic order, neglecting higher powers of $v_{EW}/f$, is just
%\beq
%V \approx - \lambda_K\, v_{EW}^2\, \tilde{K}^\dagger \tilde{K} + \lambda_K \left(\tilde{K}^\dagger \tilde{K}\right)^2
%\quad
%\text{with}
%\quad
%\lambda_K \equiv \frac{\mu_\eta^2}{2 f^2}\cos^4{\tilde\beta} + \lambda\sin^4{\tilde\beta}.
%\eeq
%The mass of the light Higgs scalar, up to $v_{EW}^2/f^2$ corrections, is therefore
%\beq
%m_h^2 \approx 2 v_{EW}^2 \lambda_K
%= v_{EW}^2 \left( \frac{\mu_\eta^2}{f^2}\cos^4{\tilde\beta} + 2\lambda\sin^4{\tilde\beta} \right)
%\label{eq:hmass}
%\eeq

\subsection{Spectrum and couplings}

The three components of each $H$ and $K$ that are not aligned with the VEV (let's denote them $h_i$ and $k_i$) are associated with the generators of the $W^\pm$ and $Z$, so they must contain exact NGB eaten by the massive gauge bosons.
Note that since we do not explicitly decompose the degrees of freedom onto the exactly broken directions once the vacuum gets misaligned, these do not correspond to the typical Goldstone parametrization.
At the level of quadratic terms that we consider here, it makes no difference, except that these $k_i$ as well as the singlet $\eta$ need to be rescaled.
One can indeed see that the kinetic term in \eqref{lagrexpand} gets multiplied by $\sin^2\theta/\theta^2$;
only the direction aligned with the VEV retains a canonical kinetic term.
So on the vacuum solution we find 
\beq
\mathcal{L} \supset - \frac{1}{2} \left( \mu_g^2 + \mu_\eta^2/\cos\theta \right)\ \left( k_i, h_i \right) \cdot
\left( \begin{array}{cc}
1 & -\cot{\tilde\beta} \\
-\cot{\tilde\beta} & \cot^2{\tilde\beta} \end{array} \right)
\cdot \left( \begin{array}{c}
k_i \\ h_i \end{array} \right)\,.
\label{GBs}
\eeq

This mass matrix has one zero eigenvalue, associated to the 3 exact GBs, while the other fields have mass
\beq
m_i^2 = \frac{\mu_g^2 + \mu_\eta^2/\cos\theta}{\sin^2{\tilde\beta}}
= \frac{\mu_K^2}{\sin^2{\tilde\beta}} + \mathcal{O}\left[ \frac{v_{EW}^2}{f^2}\right] \ ,
%= \frac{m^2}{\cos^2 \beta}\,.
\eeq
and the mixing angle is precisely $\tilde\beta$.
% $\beta$ is not exactly $\tilde\beta$ but, analogously to 2HDM scenarios, is instead
%\beq
%\tan\beta \equiv \frac{v_H}{v_K}.
%\eeq
%The two angles only differ by small corrections:
%\beq
%\frac{\tan\beta}{\tan{\tilde\beta}} = \frac{\sin\theta}{\theta} = 1 - \frac{v_{EW}^2 \cos^2{\tilde\beta}}{6\,f^2} + \mathcal{O}\left[ \frac{v_{EW}^4}{f^4}\right].
%\label{eq:betatilde}
%\eeq

The singlet $\eta$ does not mix with the other fields and has indeed mass $\mu_\eta$ up to $v_{EW}/f$ corrections:
\beq
m_\eta^2 = \frac{\mu_\eta^2}{\cos\theta}.
\eeq

Of greater relevance is the mass matrix of the `radial' scalars, $k_0$ and $h_0$, since we want to identify the lightest of the two as the observed Higgs boson:
\beq
\mathcal{L} \supset - \frac{1}{2} \left(k_0, h_0 \right)  \left( \begin{array}{cc}
\mu_g^2 \cos^2\theta + \frac{\mu_\eta^2}{\cos\theta} & 
- \frac{1}{\tan\tilde\beta} \left( \mu_g^2 \cos\theta + \mu_\eta^2 \right)
\\
- \frac{1}{\tan\tilde\beta} \left( \mu_g^2 \cos\theta + \mu_\eta^2 \right)
& 
\frac{1}{\tan^2{\tilde\beta}} \left( \mu_g^2 + \frac{\mu_\eta^2}{\cos\theta} \right) + 2\lambda v_{EW}^2 \sin^2{\tilde\beta}
\end{array} \right)\left( \begin{array}{c}
k_0 \\ h_0 \end{array} \right),
\label{higgs}
\eeq
and its eigenvalues, up to higher order terms in $v_{EW}/f$ and $\lambda$, are
\bea
m_h^2 &\approx& v_{EW}^2 \left( \frac{\mu_\eta^2}{f^2}\cos^4{\tilde\beta} + 2\lambda\sin^4{\tilde\beta} \right)
\label{eq:mh}
\\
m_H^2 &\approx& \frac{\mu_K^2}{\sin^2{\tilde\beta}}.
\eea
As for the rotation to the mass eigen-basis, the analog of the 2HDM mixing angle $\alpha$ is
\beq
\alpha = \tilde\beta - \pi/2 + \mathcal{O}\left[ \frac{v_{EW}^2}{f^2}, \lambda \right].
\label{eq:misalign}
\eeq
We are therefore close to the alignment limit and the Higgs couplings are close to SM values.

\begin{figure}[ht]
\centering
\includegraphics[width=0.7\textwidth]{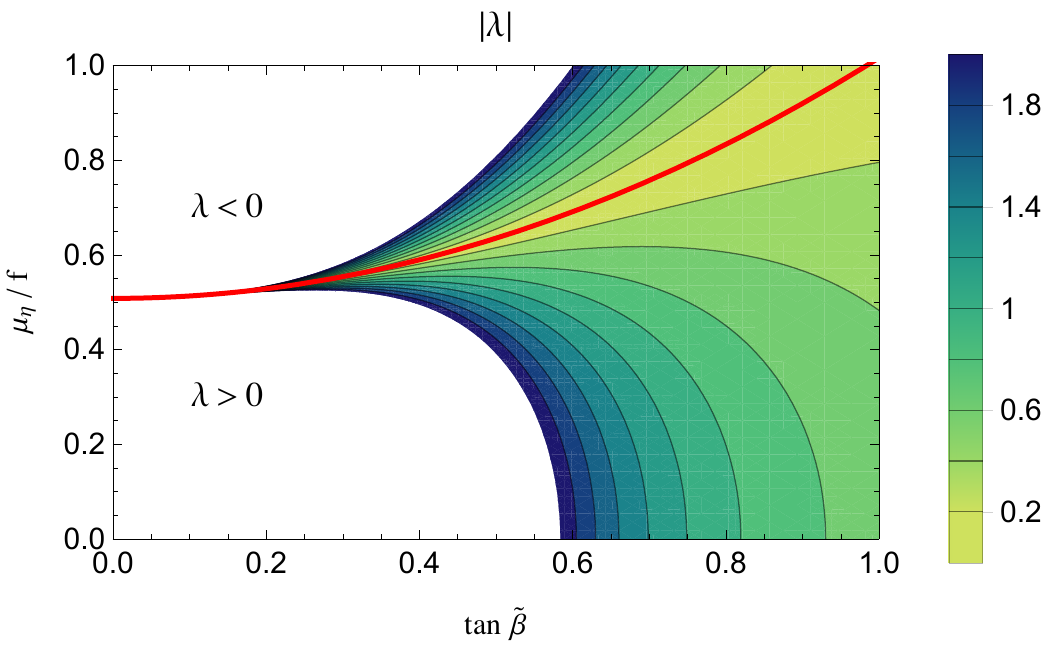}
\caption{\small{
Value of $\lambda$ required to reproduce the correct Higgs mass in the limit $f\to\infty$. The red line indicates $\lambda=0$.
}}
\label{fig:lambda}
\end{figure}

Equation \eqref{eq:mh} holds well enough when $f$ is sufficiently large, so we have the relation
\beq
\lambda \approx \frac{1}{2} \left( \frac{m_h^2}{v_{EW}^2} \frac{1}{\sin^4{\tilde\beta}} - \frac{\mu_\eta^2}{f^2} \frac{1}{\tan^4{\tilde\beta}} \right).
\eeq
It reduces to the Standard Model relation $2\lambda = m_h^2 / v_{EW}^2$ in the \mbox{$\tan\tilde\beta\to\infty$} limit, as it should;
in the opposite limit, for $\lambda$ to remain finite, $\mu_\eta$ needs to be precisely tuned to $f m_h / v_{EW}$ but the value of $\lambda$ itself is irrelevant.
Figure \ref{fig:lambda} shows how, as $\tan\tilde\beta$ is increased, $\lambda$ is required to be small unless taken negative, while a fixed value of $\mu_\eta$ is required for small $\tilde\beta$.

It is worth noting that $m_h$ does not depend on the gauge contribution to the pion masses $\mu_g$ once the electro-weak scale is fixed.
Without $\lambda$, the dark fermion masses are then necessary to reproduce the correct spectrum.
This holds even when considering the non-linearities of the potential.
Indeed, integrating out $H$, the symmetry-breaking term from the Yukawas has the same trigonometric dependence as the gauge contribution:
there cannot be an interplay between these two terms to tune the shape of the potential.
Finally the situation changes when $\mu_\eta \to 0$ (and $\lambda=0$):
one then goes to a `technicolor' limit $\theta = \pi / 2$.
In this case the position of the VEV of $K$ is fixed while $\mu_H$ is tuned against the gauge contribution to give the correct Higgs mass:
\beq
m_h^2 = \mu_H^2 \tan^2{\tilde\beta} - \mu_g^2.
\eeq

The difference compared to 2HDM regarding the couplings to gauge bosons of these scalars is the presence of extra non-linear terms.
From kinetic terms:
\bea
 &&\frac{f^2}{8}\,\mathrm{Tr}\left( D_{\mu} \Sigma^\dagger D^{\mu} \Sigma \right) = \frac{\sin^2{S}}{ S^2} \left( K^\dagger \mathcal{A}_\mu \mathcal{A}^\mu K \right)+... \\
 &&= \bigg(g^2 W_{\mu}^{+}W^{-\mu}+\frac{g^2+g^{\prime 2}}{2} Z_{\mu}Z^{\mu}\bigg) \bigg(\frac{v_{EW}^2\cos^2{\tilde\beta}}{4} + k_0 \frac{v_{EW}\cos\tilde\beta}{2} \cos\theta \bigg) +...
 \nonumber  \\
&&D_{\mu} H^\dagger D^{\mu} H =  \bigg(g^2 W_{\mu}^{+}W^{-\mu}+\frac{g^2+g^{\prime 2}}{2} Z_{\mu}Z^{\mu}\bigg)\bigg(\frac{v_{EW}^2\sin^2{\tilde\beta}}{4} + h_0 \frac{v_{EW}\sin\tilde\beta}{2} \bigg)+... \nonumber
\eea
so the expressions for the couplings contain an extra $\cos\theta$:
\bea
 \frac{g_{hVV}}{g_{hVV}^{SM}} &=&
\cos\alpha \sin\tilde\beta - \sin\alpha \cos\tilde\beta \cdot \cos\theta \\
&=& 1 - \frac{v_{EW}^2\cos^4{\tilde\beta}}{2\,f^2} + \mathcal{O}\left[ \left( \frac{v_{EW}^2}{f^2},\lambda \right)^2 \right] \nonumber
 \\
 \frac{g_{HVV}}{g_{hVV}^{SM}} &=&
 \sin\alpha \sin\tilde\beta + \cos\alpha \cos\tilde\beta \cdot \cos\theta \\
 &=& -\frac{\sin^3{2\tilde\beta}}{4} \left( \frac{1}{2} \left( 1 - \frac{\mu_\eta^2}{\mu_K^2} + \cot^2{\tilde\beta} \right) + \frac{\lambda f^2 \tan^2{\tilde\beta}}{\mu_K^2} \right) \frac{v_{EW}^2}{f^2} + \mathcal{O}\left[ \left( \frac{v_{EW}^2}{f^2},\lambda \right)^2 \right]
 \nonumber
\eea
where $ g^{SM}_{hWW}= g\, m_W$, $g^{SM}_{hZZ}=\sqrt{g^2+g^{\prime 2}}\, m_Z $.
We observe that the leading correction to the light Higgs coupling comes entirely from the `compositeness' $\cos\theta$ and the effect of the misalignment \eqref{eq:misalign} only enters at higher order.
Couplings of the singlet pion $\eta$ were computed in \cite{Cacciapaglia:2014uja}.

As we are in a Type I 2HDM, the couplings to fermions are \cite{Branco:2011iw}
\bea\label{su4:fermions}
\frac{ g_{h\bar{f}f} }{ g_{h\bar{f}f}^{SM} } &=& \frac{\cos\alpha}{\sin\tilde\beta}\ \approx\ 1 \ ,
\\
\frac{ g_{H\bar{f}f} }{ g_{h\bar{f}f}^{SM} } &=& \frac{\sin\alpha}{\sin\tilde\beta}\ \approx\ -\frac{1}{\tan\tilde\beta} \ ,
\eea
while the fundamental couplings of our elementary doublet $H$ to fermions are enhanced by $1/\sin\tilde\beta$ compared to the SM values since $v_H = v_{EW} \sin\tilde\beta$ and the fermion masses are fixed.

\section{$  \SU(5) / \SO(5) $ model}
We now repeat the same steps for the pattern of symmetry breaking $\SU(5)/\SO(5)$.
As we will see the different group theory leads to different expressions for the Higgs mass and also, contrary to the previous example, produces sizable contributions to the EDMs
of SM fermions.

This pattern of symmetry breaking is realised by an $\SO(N)$ gauge theory with 5 Weyl fermions in the vectorial representation of the gauge group.
The spontaneous symmetry breaking is driven by the dark fermion condensate $ (\Sigma_0)_{ij}=\langle \Psi_i \Psi_j \rangle $, symmetric under the exchange of its flavor indices $ (i,j=1,\dots,5)$.  The fermions transform as a lepton doublet plus its conjugate and a real singlet\footnote{Differently from the $\SU(4)/\SP(4)$ the charge assignments are compatible with unification into $\SU(5)$ multiplets. The same gauge theory models 
also predict dark matter candidates as dark baryons \cite{ACDM}.},
\be
\Psi_i=\left( Q_L ,\tilde Q_L,N \right)^\intercal=\left(\mathbf{2}_{-1/2},\mathbf{2}_{+1/2}, \mathbf{1}_0\right)^\intercal
\ee
The action is invariant under custodial symmetry acting on the vector with the $\SU(5)$ generators,
\bea\label{su5:gen}
S_L ^i = \frac{1}{2}
\left(
\begin{array}{c | c} 
\mbox{$ \mathbf{1} _2 \otimes \sigma^i $}  &   \\
\hline
  & 0 \end{array} \right)\,,
\quad 
S_R ^i = \frac{1}{2}
\left(
\begin{array}{c | c}
\mbox{$ \sigma^i \otimes \mathbf{1}_2 $ } & \\
\hline
& 0 \end{array} \right)\,,
\eea

We include dark fermions masses, $\mu_L$ for $ Q_L$ and $ \mu_N$ for $ N$, as well as Yukawa couplings, which can be incorporated in the Lagrangian as
\begin{equation}
\mathcal{L}_M + \mathcal{L}_{yukawa}=-\frac{1}{2}\Psi^\intercal M \Psi+h.c.
\end{equation}
where $ M $ is the  symmetric matrix
\begin{equation}
M =
\begin{pmatrix}
0& & & -\mu_L & \multirow{2}{*}{$\tilde y \tilde H^* $} \\
&0  &\mu_L & & \\
& \mu_L &0 & &   \multirow{2}{*}{$y  H^* $} \\
-\mu_L &  & & 0 & \\
& \multicolumn{2}{c}{   \tilde y  \tilde H^\dagger }& \multicolumn{1}{l}{   y  H^\dagger} & -\mu_N
\end{pmatrix},
\end{equation}
$ H $, as before, is the Higgs doublet, $ \tilde H ^i = \epsilon ^{ij} {H^* } ^j $, and the  fermion masses $ \mu_L $ and $ \mu_N $ are taken to be real.
%As we will see explicitly in the next subsections, the effective potential to order $ p^2 $ does depend on two independent combinations of Yukawa couplings such that, unlike in the previous model, the CP-violating phase appears to leading order.

%As these two combinations appear, let us define the following shorthand notations:
%\be
%A \equiv  y+ \tilde y ^*  , \quad B \equiv y - \tilde y^*,
%\ee
%where, as we will discuss in section \ref{su5:edm}, $ A $ is the parameter governing the possibility of having a contribution to the electric dipole moment for the SM fermions, which are only coupled to the fundamental doublet $ H $.

%We will represent the composite-Higgs field as $ K=\left(K^+, K^0\right)^\intercal$, where the upper component is the charged one, while the lower component is the neutral one. At this stage, our theory is %automatically equipped with another $ \SU(2)$ symmetry, say $ \SU(2)_R$ (simply because it acts from the right on the NGB matrix), exchanging the two NGB doublets, which, combined with $ \SU(2)_L$, %yields the usual custodial symmetry group. Explicitly, the representative matrices chosen, respectively for the algebra $ \SU(2)_L $ and $ \SU(2)_R $ are

%Note that their normalization is not the same as the corresponding $\SU(4)$ generators.

\subsection{Effective Theory}
To build the low energy effective theory for the NGB, we make the following choice for the vacuum
\begin{equation}
\Sigma_0=\left(\begin{array}{cc|c}
0 & i \sigma_2 & \\
-i \sigma_2 & 0 &\\ \hline
& & 1 \end{array}\right)
\end{equation}
The spontaneous breaking $\SU(5)\to \SO(5)$ produces 14 NGB.
Under custodial symmetry $\SU(2)_L \times \SU(2)_R$ they decompose as,
\be
14 \rightarrow (3,3)\oplus (2,2)\oplus (1,1)=\pi_a ^i \oplus (\tilde K, K) \oplus \eta
\ee
The doublet corresponds to the combination
\begin{align}\label{su5:k}
K\sim\mathbf{2}_{-1/2}\otimes\mathbf{1}_0,\quad \tilde K \sim \mathbf{2}_{1/2}\otimes \mathbf{1}_0,\,\quad
\tilde K^i =\epsilon^{ij} {K^*} ^j
\end{align}

We parametrize the NGBs as
\be\label{gbmatrix}
\Sigma \equiv \exp\left[\frac{4\,i }{f} \sum_{a=1} ^{14} X^a \Pi_a\right] \cdot \Sigma_0,
\ee
where $ X^a $ are the broken generators with trace normalized to $1/2$.
Explicitly the tri-triplet $ \pi^i _ a = (3,3) $ can be represented by means of the following block matrix
\begin{align}
\pi^i _a &=\frac{1}{2}\left( \begin{array} {c  c |c} 
\pi_0  & \pi_+& \\ 
\pi_- &  -\pi_0  \\ \hline
& & 0 \end{array}\right), \quad  \\
 \pi_0  &= \frac{1}{\sqrt{2}} \,\pi_0 ^i \, \sigma^i \,, \quad \pi_-  = \pi_- ^i \, \sigma ^i , \quad \left( \pi_- \right)^\dagger = \pi_+ \,,
\end{align}
where $ \pi_0 ^i  \in \mathbb{R} $ and $ \pi_{\pm} ^ i \in \mathbb{C}$, the subscripts $ ( a= \pm, 0 ) $ indicate the hypercharge, and the superscripts $(i=1,2,3)$ denote the components of each triplet. 

The parametrization of the doublet and that of the singlet in terms of $ 5 \times 5$ matrices are given, respectively, by
\begin{align}
\Phi &=\frac{1}{2}\left( \begin{array} {c  c |c} 
  & & \tilde K \\ 
 & & K   \\ \hline
 \tilde K ^\dagger & K^\dagger &  \end{array}\right), \quad X_\eta = \frac{1}{2 \sqrt{10}} \left( \begin{matrix} \mbox{\large{$ 1_4$}} & \\ & -4 \end{matrix} \right).
\end{align}

With the above ingredients the 2-derivative effective lagrangian reads
\bea\label{su5:lagr1}\nonumber
\mathcal{L}&=&\ \frac{f^2}{16}\, \mathrm{Tr} \left( \left(D_{\mu} \Sigma\right)^\dagger D^{\mu} \Sigma \right)
-\frac{1}{4} C_y f^3 \big(\mathrm{Tr}[M \Sigma] + h.c.\big)
-\frac{1}{2} C_g f^4 g_i^2 \mbox{Tr} [S^i\, \Sigma\, {S^i}^*\, \Sigma^\dagger] \nonumber \\
&+& \left(D_\mu H\right)^\dagger D^\mu H-\mu_H^2 H^\dagger H-\lambda \left(H^\dagger H\right)^2 \,.
\eea

%In the following subsection we will neglect any contribution to the dynamics coming from the singlet $ \eta $ and the triplets $ \pi_a ^i $, by setting the corresponding fields to zero in the Goldstone matrix \eqref{gbmatrix}. Doing so, we will be able to compute the Lagrangian \eqref{su5:lagr1} to all order in the fields, making computations simpler. Later, in section \ref{su5:full}, we will comment on the modification to the dynamics given by the introduction of the singlet and the triplets.

Focussing on the doublets the lagrangian can be written in explicit form as,
\bea\label{su5:lagr2}
\mathcal{L}&=& \ \frac{\sin^2 S}{S^2}  \left( D_\mu K\right)^\dagger D^\mu K +\frac{S^2-\sin^2 S}{2 f^2 S^4} \left(K^\dagger \partial_\mu K + \left(\partial_\mu K\right)^\dagger K \right)^2 \\ 
&+&\Big( \epsilon\,\mu_H ^2  H^\dagger K + h. c. \Big) \,\frac{\sin 2S}{2S}+\,\frac{\mu_K^2\, f^2}{2} \cos^2 S  
+(D_\mu H)^\dagger D^\mu H-\mu_H ^2 H^\dagger H-\lambda (H^\dagger H)^2 \nonumber
\eea
where we defined 
\bea
& & \mu_K ^2 = f^2  C_g \left( 3 g^2 +g'^2 \right) + 2 \,  C_y f \left( \mu_L+ \mu_N \right) \\
& & S \equiv \sqrt{\frac{2 K^\dagger K}{f^2}}, \quad  \epsilon \equiv  -\,\frac{ i\, C_y (y-\tilde{y}^*) f^2 }{\mu_H^2}\,.
\eea
As in the previous section, note that $ \epsilon $ can be made real by a redefinition of the fields $ H $ and $ K$, so we will take $ \epsilon \in \mathbb{R}^+ $, hereafter. 

The kinetic term is identical to the one found in the previous section, as expected since $ \SU(4) / \,  \SP(4)$ is a submanifold of $ \SU(5)/ \, \SO(5)$. The potential, however, has a different trigonometric dependence on $ S $. In particular, the potential generated by gauge interactions and vectorial masses both produce a $\cos^2 S$ dependence while 
the mixing with the elementary Higgs produces a potential $\sin S \cos S$. This feature is at the heart of the different prediction for the physical Higgs mass.
In this case the tuning of the electro-weak VEV can be achieved even in absence of fermion masses through the cancellation of gauge loop and heavy Higgs contribution.

As before from the second line of eq. \eqref{su5:lagr2} one finds
\begin{align}
\quad v_{EW} ^2 = v_H ^2 + f ^ 2 \sin ^2 \theta=\left(246\, \mathrm{GeV}\right)^2\,, \quad 
\text{where} \qquad
 \theta = S( \ev K )=\frac{v_K}{f} \,. \nonumber
\end{align} 
%\subsubsection{Minimization of the potential}
Minimization of the potential in \eqref{su5:lagr2} gives the following equations of motion
\bea\label{su5:vev1}
\mu_H ^2 v_H + \lambda v_H^3 &=& \epsilon\, \mu_H^2 f \sin \theta \cos \theta\,, \nonumber \\
\,\mu_H^2\, v_H \,\epsilon  \cos 2\theta &=& \mu_K ^2 \,f \, \sin  \theta \, \cos \theta\,.
\eea
These equations are quite similar to those given in \eqref{eq:vevsp4}.
%They should not depend on the details of the coset to leading order in $\theta$, and indeed match the results from the previous section when $\theta \to 0$ keeping $f\sin\theta = v_{EW}\cos\tilde\beta$ fixed.
However, the non-trivial $\theta$-dependence differs. As it is apparent from the second equation, here the gauge and fundamental fermion mass contribution, contained inside $ \mu_K ^2 $, come with the same function of $ \theta $, differently from what happened in \eqref{eq:vevsp4:2}.  We can rewrite the minimum conditions as follows
\bea\label{su5:vev}
\epsilon &=& \frac{\tan \tilde \beta}{\cos \theta} \left( 1+ \lambda \frac{v_{EW}^2}{\mu_H ^2} \sin^2 \tilde \beta \right)\approx \frac{\tan \tilde \beta}{\cos \theta} \,, \\
\label{su5:vev:2}
\mu_H ^2 \tan^2 \tilde \beta& =& \mu_K ^2 \frac{\cos^2 \theta}{\cos 2 \theta} - \lambda v_{EW}^2 \sin^2 \tilde \beta\, \tan^2 \tilde \beta \approx \mu_K ^2 \frac{\cos ^2 \theta}{\cos 2\theta}
\eea
where  $ \tan \tilde \beta $ is defined as in equation \eqref{eq:tanbtdef} and we assumed the contribution coming from elementary quartic $ \lambda $ to be much small.
As expected in the  limit $ \theta \ll 1 $ the equations above yield the tuning condition
\be
\epsilon \approx \tan \tilde \beta \approx \frac{\mu_K }{\mu_H }\,.
\ee

\begin{figure}[t]
\centering
\includegraphics[width=0.6 \textwidth]{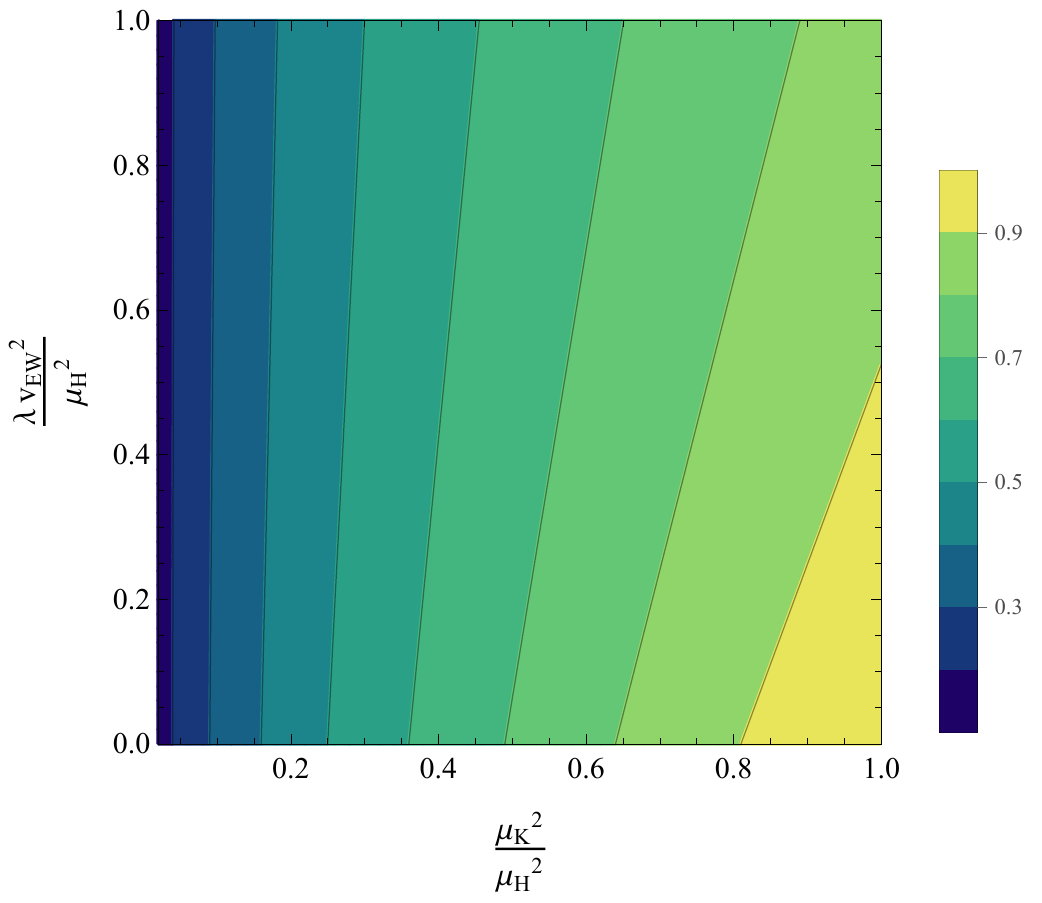}
\caption{
\small{A contour plot of $ \tan\tilde\beta $, obtained by solving equation \eqref{su5:vev:2} with $\theta=0$ as a function of $\mu_K ^2/ \mu_H ^2 $ and $ \lambda v_{EW}^2/ \mu_H ^2 $.}
}
 \label{fig:f1}  
\end{figure}

In complement to Fig.\ref{fig:lambda},
another way to understand the relation among the parameters imposed by \eqref{su5:vev:2} is to plot the value of $\tan\tilde\beta$ as a function of $\mu_K^2$ and $\lambda v_{EW}^2$ relative to $\mu_H^2$, setting $\theta=0$.
This is shown in Fig.\ref{fig:f1}.
Small values of $\tan\tilde\beta$ --- corresponding to the case where the doublet responsible for the electroweak symmetry-breaking is mostly composite --- require $\mu_H \gg \mu_K$, while the value of $\lambda$ plays no role.

%\subsubsection{Spectrum and Couplings}

Coming to the spectrum, the mass  matrix of the CP even neutral states $ (k_0, h_0 ) $ 
%after removing the parameters $ \epsilon $ and $ \mu_H $ by use of the VEV equations \eqref{su5:vev}, 
is given by
\bea\label{su5:mass}
 \left(k_0, h_0 \right)
\left(
\begin{matrix}
 \frac{\mu_K ^2}{ \cos 2 \theta }& -\frac{\mu_K ^2}{\tan \tilde \beta} \cos  \theta \\
-\frac{\mu_K ^2}{\tan \tilde \beta} \cos  \theta &\quad \frac{\mu_K^2}{\tan^2 \tilde \beta}\frac{\cos ^2 \theta}{\cos 2\theta}+2 \lambda v_{EW}^2 \sin ^2 \tilde \beta  \end{matrix} \right)\left( \begin{array}{c}
k_0 \\ h_0 \end{array} \right)\,.
\eea
To first order in $ \lambda$ and up to corrections of order $\left( v_{EW}^4/f^4 \right)$ the eigenvalues are
\bea\label{su5:higgsmass}
m_h^2 &\approx& 4 \,\frac{v_{EW}^2}{f^2} \mu_K ^2 \cos ^4 \tilde \beta+ 2\, \lambda \, v_{EW} ^2 \sin ^4 \tilde \beta \,,  \\
m_H ^2 & \approx &  \frac{\mu_K ^2}{\sin^2 \tilde \beta} +\, 2  \lambda\,   v_{EW}^2\cos^2 \tilde \beta \sin ^2 \tilde \beta \,. \nonumber 
\eea
The small eigenvalue, $ m_h $, is the one associated with the standard model Higgs boson, while $ m_H $  is the mass of its heavy partner.  
Importantly, contrary to what occurred in the previous section, $ \mu_K ^2$ contains a contribution from gauge loops allowing to tune the Higgs even without quark masses. 
In this limit using the estimate of gauge loops from QCD data one finds \cite{Georgi:1984ef},
\begin{equation}
m_h^{gauge}\approx 150\,\sqrt\frac3 {N_D}\, \rm GeV
\end{equation}

Finally, the relation between the angle $ \tilde \beta $ and the angle $ \alpha $, the latter corresponding to the rotation needed to diagonalize the mass matrix \eqref{su5:mass}, is exactly that given in equation \eqref{eq:misalign},  up to $ \mathcal O  \left[ \frac{ v_{EW}^2}{f^2} , \lambda  \right] $ . The couplings of the light Higgs to a pair of gauge bosons are equal to those computed in the previous section, while the one of the heavy Higgs is
\bea
\frac{g_{H V V}}{g_{h V V}^{SM}}\approx-\frac{\sin^3 {2\tilde\beta}}{4} \left( -1 + \cot^2{\tilde\beta} + \frac{\lambda f^2 \tan^2{\tilde\beta}}{\mu_K^2} \right) \frac{v_{EW}^2}{f^2} \,.
\eea
The couplings to fermions are equal to those given in \eqref{su4:fermions}.

The remaining scalars take mass from the Lagrangian
\bea
\mathcal{L}_{mass}&=&-\,\frac{\mu_K^2\cos^2 \theta}{2\cos 2 \theta} 
\left(\begin{array}{ccc}
k_i &,& h_i \end{array} \right) 
\cdot 
\left(
\begin{matrix}
1 & - \cot \tilde \beta\\
- \cot  \tilde \beta & \cot^2 \tilde \beta \end{matrix} \right)
\cdot
\left(\begin{array}{c}
k_i \\
 h_i \end{array} \right)  \,,
\eea
leading to
\bea
m_i &=& \mu_K ^2 \, \frac{\cos^2 \theta}{\cos 2 \theta} \frac{1}{\sin^2 \tilde \beta}  \approx  \frac{\mu_K ^2}{\sin^2 \tilde \beta} + \mathcal{O} \left(\frac{ v_{EW}^2}{f^2}\right)\,,
\eea
where we have made explicit use of equations \eqref{su5:vev},  \eqref{vewdef}.

\subsection{Electric Dipole Moments}\label{su5:full}

\begin{figure}
\centering
\includegraphics[width=0.5 \textwidth]{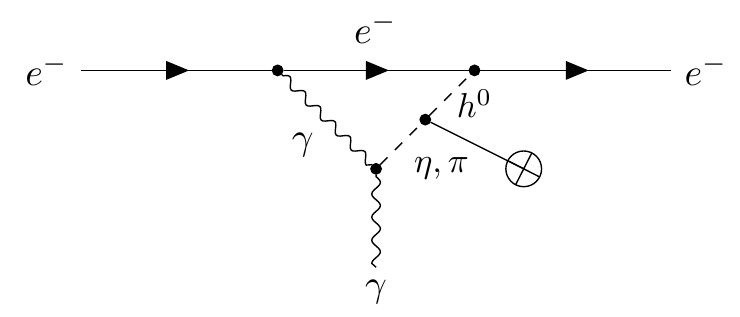}
\caption{\small{Feynman diagram contributing to the electric dipole moment of the electron. The $ \eta $ and $ \pi $ NGB couple to photons through anomalies, while the crossed circle represents the insertion of the Higgs VEV.}}
\label{fig:edm} 
\end{figure}

 For complex Yukawa couplings the models contain a new CP violating phase.
As in Ref. \cite{Antipin:2015jia} this phase will in general induce EDMs for the SM particles\footnote{For $\SU(4)/\SP(4)$  the leading order Lagrangian is accidentally CP invariant, see appendix. As a consequence, no EDMs are generated to leading order. For this reason we focus on $\SU(5)/\SO(5)$.}. In the microscopic theory  these originate from 2-loop Barr-Zee diagrams with the elementary fermions running in the loop \cite{Barr:1990vd}. To estimate the EDMs in the confined regime we turn to effective theory, where the singlets and triplets have anomalous couplings to the photon. Including CP violating couplings from the scalar potential an EDM can be generated through the diagrams of the kind depicted in 
figure \ref{fig:edm}. 

For $\SU(5)/SO(5)$ the anomalous couplings  read \cite{Ferretti:2016upr}
\begin{equation}
{\cal L}_{anomaly}= \frac{N_D}{16 \pi^2 f} \left[\sqrt{\frac{2}{5}}\, \eta \cdot \left( g^2 W_{\mu \nu}^i \tilde W^{\mu \nu, i}   +g'^2 B_{\mu \nu} \tilde B^{\mu \nu} \right)+ \sqrt{2}\, g\cdot g' \, \pi^i _0 \cdot W_{\mu \nu}^i \tilde B^{\mu \nu}  		\right]
\end{equation}
where $ \tilde V^{\rho \sigma} = \frac{1}{2} \varepsilon^{ \rho \sigma \mu \nu} V_{\mu \nu} $.
%In this section, since we consider the contribution to the dynamics given by the singlet and the triplets, we need to do an expansion of Lagrangian \eqref{su5:lagr1}, in terms of $ \Pi /f $, and %then it is useful to have a perturbative expansion for $ v_{EW} $, which can be written as
%\be\label{ew}
%v_{EW} ^2 = v_H ^2 + v_K ^2 + \mathcal{O}\, \left(  \frac{v_{K} ^4}{f ^4} \right),
%\ee
%which naturally defines the angle $ \beta $ for the usual two-Higgs doublet model (2HDM) as $\tan \beta \equiv \frac{v_H}{v_K} $. By the definitions, the relation between $ \tan \beta $ and $ \tan \tilde \beta $ is given by
%\begin{align}\label{eq:betatildebeta}
%&\tan \beta =\frac{\sin\theta}{ \theta} \tan \tilde \beta = \tan \tilde \beta \left( 1 - \frac{v_{EW}^2\, \cos ^2 \tilde \beta }{6 f^2} + \mathcal{O}\left( \frac{v_{EW} ^4}{f^4}\right) \right).
%\end{align}
%Thus, we have that $ \tan \beta \approx \tan \tilde \beta $ up to corrections of order $ v_{EW}^2/f^2$.
For the potential it will be sufficient for our purposes to expand  \eqref{su5:lagr1}, up to quartic order in $ \frac{\Pi}{f} $ around the origin, 
\bea \label{su5:pot}
V&\approx& \mu_K ^2\, K^\dagger K \ + \ \mu_H ^2\, H^\dagger H \ +\ \lambda\, (H^\dagger H) ^2    \\
 &-&  \epsilon\, \mu_H ^2 \left( \, H^\dagger K +  K^\dagger H \right)\left( 1- \frac{4}{3 f^2} \left( K^{\dagger} K \right) \right)-\lambda_K \left( K^\dagger K \right) ^2 	\nonumber  \\
 &+&  \frac{3}{ \sqrt{10}} \, C_y f \, \eta \,\left( A\, H^\dagger K +A^* K ^\dagger H \right)+   \ \frac{m_\eta ^2}{2} \,\eta ^2 \ + \ \frac{m_{\pi_0 }^2}{2}\, (\pi_0 ^i  )^2
\ +  \ m_{\pi_+ }^2\, \pi_+ ^i \cdot \pi_ - ^i  \nonumber \\
&+&  C_y \, A \,   f \left( - H^\dagger \sigma^i \tilde K \pi_- ^i +\frac{1}{\sqrt{2}}H^\dagger \sigma^i K \pi_0 ^i \right) +\mathrm{h.c.} \,, 	\nonumber 
\eea
where
\begin{align}\label{param}
 \lambda_K  \equiv & \,\frac{2}{3}\,\frac{\mu_K ^2}{f^2} , \quad  m_\eta ^2  \equiv \frac{ 4}{5}\,C_y \, f \left( \mu_L + 4 \mu_N\right) , \quad 
  m_{\pi ^0}^2 \equiv   \ 4\, f^2 \left( \frac{C_y\, \mu_L}{f} + 2\, C_g \, g^2  \right), \quad \\
 m_{\pi^+}^2 \equiv & \,  4\,f^2 \left( \frac{C_y\, \mu_L}{f} +
 C_g \left( 2\,g^2 + g'^2 \right)\right)\,,~~~~~~~~~~~~~~~~A\equiv y+\tilde{y}^*\,,~~~~~~~~~~B\equiv y-\tilde{y}^*.  \nonumber
\end{align}

Using this expansion the same results of the previous section can be recovered up to $ \mathcal{O} \left(v_{EW}^4/f^4 \right)$ corrections.
We can also extract the VEV of singlets of triplets
\begin{align}\label{su5:vevs}
\ev{\eta}&=-\frac{6}{\sqrt{10}} \frac{ C_y^2\,f^3\, \mathrm{Im}\left(y \tilde y \right)\,v_K ^2}{\, m_\eta ^2\,\mu_H^2} \,,~~~~~~~~~\ev{\pi^3 _0 }=\frac{\sqrt{2}\, C_y^2\,f^3 \,\mathrm{Im}(y \tilde y)\,v_K^2}{m_{\pi ^0 }^2\, \mu_H ^2} \,,  \nonumber \\ 
\ev{\pi_+ ^0} & \equiv \frac{\ev{\pi_+ ^1} - i \ev{\pi_+ ^2 } }{\sqrt{2}} = \frac{C_y^2 f^3 v_K^2}{\sqrt{2}\, m_{\pi^+} ^2\, \mu_H ^2 }\, [2 \mathrm{Im} (y \tilde y) + i \left( \lvert y \rvert ^2 - \lvert \tilde y \rvert ^2 \right) ]
\end{align}				
%Consistently the triplet VEVs only contribute to higher order in $ v_{EW}^2/f^2$, being proportional to $ v_K ^2 $.

%\subsubsection{Electric Dipole Moments} \label{su5:edm}
The contribution to  EDMs can be computed as in \cite{Antipin:2015jia}. We first integrate out the elementary doublet $ H $ 
at tree level in \eqref{su5:pot}, obtaining  the following relevant Lagrangian
\bea
\delta \mathcal{L}^{EDM}&=& -\ \frac{m_{\pi _0}^2}{2}\left( \pi_0 ^3 \right)^2 -\, \frac{m_\eta ^2}{2} \eta^2 +\frac{4 \, C_y^2\,f^3\, \mathrm{Im}\left( y \tilde y \right)}{\mu_H ^2} \left( \frac{1}{\sqrt{2}} \, K^\dagger \sigma ^i K \, \pi_0 ^i  + \frac{3}{\sqrt{10}} \, \eta \, K^\dagger K \right) \nonumber \\
&+&\mathcal{L}_{anomaly} \,.
\eea
Next we integrate out $ \pi_0 ^i $ and $ \eta $ assuming them to be heavier than the physical Higgs,
\bea
\delta \mathcal{L}^{EDM}_{eff} &=&\frac{ N_D \, C_y^2\,f^2 \,\mathrm{Im}(y \tilde y)}{5\,\pi^2 \mu_H^2\, m_\eta ^2 \, m_{\pi_0}^2 } \left[ 3 m_{\pi_0}^2\, K^\dagger K \left(g^2 W_{\mu \nu}^i \tilde W^{\mu \nu, i} +g'^2 B_{\mu \nu} \tilde B^{\mu \nu} \right)+ 5\, m_\eta ^2\,  K^\dagger \sigma^i K \, g g'\,W_{\mu \nu}^i \tilde B^{\mu \nu} \right]
 \nonumber \\
\eea
from which we can extract the coupling of the Higgs to photons, which reads ($ \lambda \ll 1 $)
\bea \label{edm}
\mathcal{L}^{EDM}_{\gamma \gamma} &=&\frac{e^2 N_D}{40\, \pi ^2 } \frac{ \mathrm{Im}(y \tilde y) \,C_y^2\,f^2 \left(6 m_{\pi_0} ^2\,  - 5 m_\eta ^2 \right) \, \tan^2 \beta}{\mu_K ^2 m_{\pi_0}^2 m_\eta ^2} \, F \tilde F \, k_0 ^\dagger\, k_0 \equiv \frac{c_K}{\Lambda^2}F \tilde F \, k_0 ^\dagger\, k_0\, ,  
\eea
and in the last equality we defined
\bea
\qquad  \frac{c_K}{\Lambda^2} \equiv \frac{e^2\, C_y^2 f^2 \,N_D  \mathrm{Im}(y \tilde y)}{40\,\mu_K ^2  \pi^2} \times\frac{ \left( 6\, m_{\pi_0}^2 - 5\,m_\eta^2 \right) }{m_{\pi_0}^2 m_\eta^2}\times  \tan^2 \beta \,.  
\eea
The electric dipole moment can be estimated in the effective theory from the one-loop diagrams generated through the effective operator in \eqref{edm}. 
For the electron this gives \cite{Lue:1996pr,splitsusy2}
\be\label{edmformula}
d_e \approx\frac{e\,m_e\, c_K}{4\, \pi^2 \Lambda^2} \, \log \frac{m_\Pi^2}{m_h^2}.
\ee
where the factor $m_\Pi$ in the $\log$ is an average of pion masses that cuts off the logarithmically divergent integral obtained from the effective operator (\ref{edm}).
%Note that we defined $\Lambda$, the combination that enters the logarithm,  to depend on the physical pion masses as in the effective theory
%these cut-off the logarithmically divergent integral \cite{Lue:1996pr}.
Restoring the  powers of $ \hbar $ and $ c $ in the previous formula, we find the following estimate for the electric dipole moment of the electron
\be
d_e \thickapprox 10^{-26} \, \mathrm{e}\cdot \mathrm{cm} \times \mathrm {Im} (y \tilde y) \times \frac{N_D}{3} \times \left( \frac{\mathrm{TeV}}{m_{\pi,\eta}} \right) ^4 \times \Big( \frac{m_\rho}{\mathrm{TeV}}\Big)^2 \times \tan ^2 \beta \,.
\ee
This prediction should be compared with the current experimental limit \cite{Baron:2013eja}, namely $ \lvert d_e \rvert < 8.7 \times 10^{-29} \mathrm{e}\cdot \mathrm{cm} $, at $ 90 \% $ confidence level, pointing towards the fact that new composite particles could be detected at the TeV scale, provided that the CP-violating phases are of order one.

\subsection{Custodial Symmetry breaking}
\label{sec:custodialSO}
Finally, we discuss the violation of custodial symmetry in the \SU(5)/\SO(5) model. For $y\ne  \tilde{y}$ custodial symmetry is broken by the Yukawas.
While the VEV of the doublet produces no tree level correction to $T$ the vevs listed of the triplets in eq.\eqref{su5:vevs}  break in general custodial symmetry producing the following contribution to the $ T $ parameter
\be \label{su5:tparam}
T =  \frac{8}{\alpha_{em}} \, \left(\frac{ C_y^2\,f^3 }{m_{\pi ^+ }^2 \, m_{\pi^0}^2 \, \mu_H ^2} \right)^2 \,v_{EW} ^2 \, c_{\beta}^4  \left[ \mathrm{Im}(y \tilde y ) ^2\left( m_{\pi ^+ }^4 - \, m_{\pi ^0 }^4\right) +\frac{m_{\pi^0}^4}{4} (| y | ^2 - | \tilde y | ^2 ) ^2  \right]
\ee
where $ \alpha_{em} = e^2/4\pi $. The first term on the right hand side of eq.\eqref{su5:tparam} vanishes when $ m_{\pi^+} = m_{\pi^0} $, namely when $ g'=0 $, thus is essentially a hypercharge effect, combined with the complex phase $ \mathrm{Im}(y \yt)$; the second term, on the contrary, is a purely yukawa effect, depending only on the difference between the two yukawas $ y $ and $ \yt$, and is present even when the theory features no CP violation. In the $\SU(4)/\mathrm{Sp}(4)$ model there is no such effect because of the absence of the triplets, and corrections to the $ T $ parameter are generated at the loop level.

\section{Relaxing the Composite Higgs}
\label{sec:relaxion}

The composite Higgs models that have been presented, feature an elementary Higgs doublet, necessary to  induce the vacuum misalignment
of the composite Higgs that dominantly breaks the electro-weak symmetry. The presence of the elementary scalar reintroduces the
hierarchy problem, apparently off-setting the very motivation of composite Higgs models.
\
In this section we sketch  how to realize the relaxation of the electro-weak scale  \cite{relaxion} for the composite Higgs.
In \cite{Antipin:2015jia}  we discussed the relaxation in the limit where the physical Higgs is mostly elementary.
We here turn to the composite regime. This is the composite analog of the supersymmetric relaxion mechanism studied in \cite{Hardy:2015laa,susyrelaxion,gherghetta}. 

Following \cite{relaxion} we will assume that the elementary Higgs mass scans in the early universe due to the evolution of
the relaxion scalar field $\phi$,
\begin{equation}
V(\phi,H)= (\Lambda^2 - g \Lambda \phi) |H|^2+ g \Lambda^3 \phi+\dots
\label{relaxpot}
\end{equation}
This potential is technically natural, sending $g\to 0$ invariance under shift of $\phi$ is recovered.
In addition the relaxion has an axion like coupling to the dark gluons,
\begin{equation}
\frac 1 {32\pi^2} \frac {\phi}{f_\phi} \tilde{\mathcal G}_{\mu\nu}^a {\mathcal G}^{a,\mu\nu} \ .
\end{equation}
where $f_\phi$ is a decay constant (not to be confused with $f$) of the order of the cut-off of the theory $\Lambda$.
This term can be eliminated through a chiral rotation of the fermions so that $\phi$ appears in the masses and Yukawas of the fermions
and we will work in this basis in what follows, see \cite{Antipin:2015jia} for more details. We neglect derivative couplings.

%The vacuum energy is,
%\begin{equation}
%E(\theta_H)\approx  -{\rm Re}[4 m_L+ 2 m_N] g_\rho f_\pi^3  +  2\frac{g_\rho^2 f_\pi^4}{m_{H_2}^2}\left[|y|^2+|\yt|^2 - 2 {\rm Re}[y\yt] \right] |K|^2 + {\cal O}(K^4)
%\label{relaxionpot}
%\end{equation}

%Formally the only difference from \cite{halfcomposite} is that also $m_H$ is function of $\phi$. 
As we discussed in section \ref{sec:models} the tuning of the electro-weak VEV 
requires that the mass matrix (\ref{mass22}) for the elementary and composite Higgs  has a small negative eigenvalue. 
This happens for,
\begin{equation}
m_K^2 \approx |\epsilon|^2 m_H^2
\label{relaxtuning}  
\end{equation}
To see  how the tuned  electro-weak scale is selected assume that $m_H^2(\phi)$ in (\ref{relaxpot}) is initially  positive. 
For $m_H^2(\phi) \gg m_K^2$  we can integrate out $H$ and obtain an effective potential for $K$,
\begin{equation}
V_K= \left(m_K^2 - |\epsilon(\phi)|^2  m_H^2(\phi)\right) |K|^2+\dots
\label{relaxVK}
\end{equation}
where $\epsilon$ depends on $\phi$ through the Yukawa couplings and mass of $H$. 
During the evolution of $\phi$ the mass of $K$ scans crossing zero when the condition
(\ref{relaxtuning}) is satisfied. Afterwards $K$  acquires a VEV that in turn generates a barrier for  $\phi$ 
that can terminate its evolution.  In light of eq. (\ref{relaxtuning}) the height of the Higgs dependent barrier is,
\begin{equation}
B_H \sim m_K^2 v^2
\end{equation}
The derivative of the barrier is given by,
\begin{equation}
\frac {\partial}{\partial \phi} V_K= \frac {\partial}{\partial \phi}\left[|\epsilon|^2 m_H^2(\phi)\right] v^2 \sim \left(\frac 1 {f_\phi} |\epsilon|^2  m_H^2(\phi) + |\epsilon|^2 g \Lambda\right) v^2
\end{equation}
Equating this with the derivative of (\ref{relaxpot}) the first local minima of the $\phi$ potential 
will be located at\footnote{The first term in eq. (\ref{relaxtuning}) dominates for $m_H^2 > g \Lambda f_\phi$ so that $m_H^2 \Lambda^2> m_K^2 v^2$.}
\begin{equation}
g \Lambda^3 f_\phi \sim m_K^2 v^2 
\end{equation}
This equation implies that choosing $g \ll 1$ a large hierarchy $v \ll \Lambda$ can be generated \cite{relaxion}.

For the relaxion mechanism to work the Higgs dependent barrier of $\phi$ must dominate other barriers. 
In the effective theory a Higgs independent barrier  is obtained by closing the $K$ loop in eq. (\ref{relaxVK}).
From this it follows that the effective cut-off  of these loops (corresponding to the scale of compositeness) must be below $4\pi v$.
Since $\phi$ scans the $\theta$ angle of the dark gluons we also need the $\theta$ dependence of the vacuum energy to be subleading. 
For example for  $m_L > m_N$ the vacuum energy scales as,
\begin{equation}
E(\theta_D)\sim m_N g_\rho f^3 \cos (\theta_D -\theta_0)
\label{eq:E0}
\end{equation}
If the fermion masses
dominate $m_K$ the Higgs dependent term dominates the energy for,
\begin{equation}
\frac {m_L}{m_N}> \frac {f^2}{v^2}
\end{equation}
Note that for degenerate masses the Higgs dependent barrier is always smaller than eq. (\ref{eq:E0})
and the classical rolling $\phi$ cannot be stopped.
 
\subsection{High Scale Relaxation} 

The relaxion mechanism described above requires the masses of the new fermions to be close to the electro-weak scale.
While this is technically natural the coincidence of scales is not satisfactory.
 In \cite{pomarol} a mechanism to push the new physics to much higher scales was described. This requires the existence of
a second rolling field, $\sigma$, scanning the barrier height in such a way that it becomes much smaller than the 
typical value. 

This can also be implemented in our framework  allowing to push the scale $f$ to much higher values. 
The dynamics generating the relaxion potential is identical to the one discussed  in the appendix A of \cite{pomarol}, 
but in a different region of parameters.  The field $\sigma$ scans the vectorial masses of the fermions,
\begin{equation}
(m_L^0 + g_L^\sigma \sigma+ g_L^\phi \phi) L L^c+ (m_N^0 + g_N^\sigma \sigma+ g_N^\phi \phi) N N^c
\end{equation}
By assumption $\sigma$ does not couple to the topological density and to the Higgs, a choice not dictated by the symmetries but radiatively stable. 
Assuming for simplicity that $m_N$ is the lightest fermion one finds, 
\begin{equation}
E(\theta_D)\sim (m_N^0 + g_N^\sigma \sigma+ g_N^\phi \phi) g_\rho f^3 \cos(\theta_D-\theta_0)
\end{equation}
In the early universe $\phi$ is initially trapped in a deep minimum of the potential and $\sigma$ scans. Generically $E(\theta_D)$ will cross zero 
and at that point $\phi$ will start evolving as well. 

Higher order effects should also be taken into account. If the $m_L$ is lighter than the confinement scale as we are assuming to construct a composite Higgs 
these arise as higher order corrections in the chiral lagrangian of the composite Higgs. For example terms with 2 mass insertions scale as,
\begin{equation}
\Delta E \sim  m^2 f^2 U^2
\end{equation}
and cannot be cancelled with the mechanism above due to the different trigonometric dependence.
This barrier is smaller than the Higgs dependent one for,
\begin{equation}
\frac {m_N^2}{m_K^2}< \frac {v^2}{f^2}
\end{equation}
Using this mechanism  $f$ can be much larger than the TeV. We leave a detailed analysis to future work.

\section{Conclusions}
\label{sec:conclusions}

In this paper we have constructed a simple UV completion of composite Higgs models. 
At first sight a disheartening  aspect of our construction is the presence of an elementary doublet needed to trigger electro-weak symmetry.
An elementary scalar however is most welcome  to construct a viable theory of flavor.
Indeed from this point of view the benefits of the elementary Higgs are as nice as in the SM. 
The only breaking of the SM flavor symmetries  is due to the SM Yukawa couplings so that the theory automatically respects minimal flavor violation; 
contributions to electro-weak parameters are well under control. 
Moreover these theories predict new states and deviation from the SM that could be visible at LHC if the dynamical scale is around TeV.

We have shown the conditions under which a successful phenomenological model is obtained. In particular the
$T$ parameter requires a custodially preserving vacuum that can be more easily realized in $\SO(N)$ and $\SP(N)$ gauge theories.
In $\SU(N)$ theories further assumptions have to be made due to the presence of several composite Higgs bosons.
In all cases the Higgs mass is predicted in terms of the parameters of the UV Lagrangian. When the Higgs mass is dominated 
by SM gauge loops this is naturally of the right size to reproduce 125 GeV. 

The models described here interpolate between elementary and composite Higgs models. In the composite regime the
dynamics shares many of the features of composite Higgs models discussed in the literature concerning the Higgs and spin-1 resonances. 
SM fermions remain elementary so that they have properties close to the SM. Sizable EDMs however may be generated by the CP
violating phases of the new sector. One interesting consequence of our construction is that the Higgs cannot be arbitrarily composite if the theory must be valid up to a high scale. Indeed in order to reproduce the known fermion masses in the SM the Yukawa coupling with the elementary Higgs must be larger than in the SM. 
Absence of Landau poles then translates into a lower bound on the mixing with the elementary Higgs.

In the last part of the paper we attempted to connect the elementary Higgs required to induce electro-weak symmetry breaking with naturalness. 
In order to achieve this we showed how to implement the relaxion mechanism in our framework. In the simplest scenario $f\sim$ TeV
but it is also possible to push the compositeness scale higher deviating from the minimal scenario with multiple  relaxion fields.

\subsubsection*{Acknowledgments}
DB and MR are supported by the MIUR-FIRB grant RBFR12H1MW.  OA thanks the Mainz Institute for Theoretical Physics (MITP) for hospitality during
part of this work. We wish to thank Andrea Mitridate, Alessandro Strumia, Andrea Tesi, Mattia Crescioli, Francesco Ciumei and Elena Vigiani 
for discussions on related subjects.

\appendix
\section{The mass matrix and CP violation}
\label{app:mass}

The UV Lagrangian of the models in section 3 and 4 contains one physical CP violating phase. However in the two models 
this has very different effect leading to sizable EDMs for the electron in the second case and smaller one in the first.
In this appendix we wish to show how this feature emerges from the different symmetry structure of the two models.

First let us comment on the allowed structure of fermion bilinears, and so in particular of Yukawa terms we are interested in.
Thinking of a basis where all the quarks are left-handed Weyl spinors, the fermion bilinears one can form have their Dirac indices contracted anti-symmetrically.
Their dark color indices are contracted \mbox{(anti-)symmetrically} if they are in a \mbox{(pseudo-)real} representation of the underlying gauge group.
This implies that the flavor structure needs to be anti-symmetric or symmetric if the breaking is respectively $\SU(N)/\SP(N)$ or $\SU(N)/\SO(N)$;
and this in turn means that the mass/Yukawa matrix needs to decompose onto the broken generators as we will now argue.

The broken generators $X$ and the unbroken ones $S$ satisfy the relations
\beq
X \Sigma_0 =  \Sigma_0 X^\intercal,
\qquad
S \Sigma_0 = -\Sigma_0 S^\intercal.
\eeq
Since the vacuum $\Sigma_0$ preserved by $\SP(N)$ is anti-symmetric, the products $X\Sigma_0$ (or equivalently $\Sigma_0^\dagger X$) are going to be anti-symmetric as well, while the ones involving the unbroken generators are symmetric;
the converse is true for $\SO(N)$.

The $n(n\mp 1)/2-1$ structures $\Sigma_0^\dagger X$, supplemented by the vacuum itself, form a complete basis of anti-symmetric or symmetric matrices when the stability group is $\SP(N)$ or $\SO(N)$.
The matrix $M$ can thus be written as
\beq
M = \Sigma_0^\dagger \tilde{M}
\eeq
where $\tilde{M}$ is a complex linear combination of $(\mathbb{1},X)$, and the term appearing in the lowest order chiral Lagrangian becomes
\beq
\mathrm{Tr}[M \Sigma] = \mathrm{Tr}[\Sigma_0^\dagger\, \tilde{M} e^{i \Pi}\, \Sigma_0] = \mathrm{Tr}[\tilde{M} e^{i \Pi}].
\eeq
In other words the Yukawa couplings of the elementary Higgs doublet can be written as a combination of broken generators, precisely the ones corresponding to the composite doublet $K$.
Note that this argument holds just as well for $\SU(N)\times\SU(N)/\SU(N)$ cosets.

Next, we already pointed out in the main discussion how one CP-violating phase is expected.
After making the quark masses real, the remaining re-phasing freedom is
\beq
y \to e^{i \delta} y,
\qquad
\yt \to e^{-i \delta} \yt;
\eeq
so any one combination of $y$ and $\yt$ can be made real, however only a single given combination at a time, leaving a non-trivial relative phase between the Yukawa couplings.
Having one elementary Higgs doublet and one composite, there is a unique linear mixing operator, $H^\dagger K$, the coefficient of which can thus always be made real.
CP violation therefore needs to come from further terms.
In particular we are interested in terms of the form $H^\dagger K \eta$:
since the $\eta$ couples to electro-weak anomalies, it would then lead to EDM effects.

Such trilinear couplings come from the trace $\mathrm{Tr}[M \Sigma]$.
As we have just shown, $M$ also, not just $\Sigma$, decomposes onto the broken generators $X$;
this means the relevant trace --- picking up one $H$, one $K$, and one $\eta$ --- is of the form $\mathrm{Tr}[X X X]$.
For a symmetric coset, the product of two broken generators is
\beq
2\, X^a X^b = \{ X^a, X^b \} + [ X^a, X^b ] = \frac{1}{n} \delta^{ab} \mathbb{1} + d^{abc} X^c + i\,f^{abc} S^c.
\eeq
Taking the trace of three broken generators thus selects the symmetric structure constants $d^{abc}$.

The $\SU(4)/\SP(4)$ coset is special in that respect:
these $d$ coefficients over the broken coset are all zero.
And this is why the expected CP violation does not appear in the lowest order chiral Lagrangian, as we have observed when presenting \eqref{lagrexpand}, contrary to the situation in the other model.

To find terms involving the CP-violating phase in the first model, we need to go beyond the Lagrangian \eqref{lagr} and consider the interplay between the quark mass terms and Yukawa couplings in higher order terms.

The next invariants we have are
\beq
 \mbox{Tr} [ M \Sigma M \Sigma]  \qquad   \text{and} \qquad \mbox{Tr} [ M \Sigma]^2
\label{LEDM}
\eeq
the terms which after expansion leads to linear couplings of $\eta$ of the form we seek are
\bea
\mathcal{L_{\eta}}
&\sim& 4\sqrt{2}\, \eta\, \frac{\sin^2{S}}{ S^2}
\Bigl(
\mathrm{Re}(H^\dagger K) \mathrm{Re}(\# (y+\yt)(\mu_L - \mu_R) )
- \mathrm{Im}(H^\dagger K) \mathrm{Im}(\# (y-\yt)(\mu_L - \mu_R) )
\Bigr) \nonumber\\
&&- 2\sqrt{2}\,f\, \eta\, \frac{\sin{2S}}{S}\, \mathrm{Im}(\# (\mu_L^2 - \mu_R^2) )
\eea
where $\#$ designates the coefficient of the operators \eqref{LEDM} in the chiral Lagrangian.
Even if this latter coefficient is real (also setting the quark masses to be real) one is left with
\beq
\#\, 4\sqrt{2}\, \eta\, \frac{\sin^2{S}}{ S^2}\, (\mu_L - \mu_R)\, \mathrm{Re}\left( (y + \yt^*) H^\dagger K \right)
\eeq
which is the expected term, proportional to the other (in general complex) Yukawa combination, $y+\yt^*$, and to the $\SP(4)$-breaking mass $\mu_L - \mu_R$.

The reason why only the $\SP(4)$-breaking mass appears can be understood similarly to the absence of a term of interest in the leading-order Lagrangian, using the argument of appendix \ref{app:mass}:
the $\SP(4)$-breaking mass is the term in the mass matrix $\tilde{M}$ proportional to a broken generator while the symmetry-preserving one is proportional to the identity.
One sees immediately that the traces \eqref{LEDM} are zero if they involve an identity matrix instead of a broken generator for the quark mass contribution.

So any EDM effect depends on additional unknown higher order coefficients and suffer from and extra $\mu/f$ suppression, which makes its study of limited interest in the $\SU(4)/\SP(4)$ model.
This is even reinforced by the fact that the anomalous coupling of $\eta$ to photons is zero at leading order \cite{Arbey:2015exa};
one would involve instead a $Z$ boson in the loop via $g_{\eta Z\gamma}$, leading to yet another suppression due to its smaller coupling to the electron.

\bibliographystyle{JHEP}
\footnotesize
\bibliography{references}

\end{document}